%% file: paper8.tex
\documentclass[twocolumn, aps, pra, superscriptaddress, floatfix, nofootinbib]{revtex4-2}

\input{packages}
\input{definitions}

\begin{document}

\title{Direct Equivalence between Dynamics of Quantum Walks and Coupled Classical Oscillators}

\author{Lilith Zschetzsche\,\orcidlink{0009-0002-9531-4039}}
\email{lilith.zschetzsche@univie.ac.at}
\affiliation{University of Vienna, Faculty of Physics, Boltzmanngasse 5, 1090 Wien, Austria}

\author{Refik Mansuroglu\,\orcidlink{0000-0001-7352-513X}}
\email{refik.mansuroglu@univie.ac.at}
\affiliation{University of Vienna, Faculty of Physics, Boltzmanngasse 5, 1090 Wien, Austria}

\author{András Molnár\,\orcidlink{0000-0001-5144-438X}}
\affiliation{\mbox{University of Vienna, Faculty of Mathematics, Oskar-Morgenstern-Platz 1, 1090 Vienna, Austria}}

\author{Norbert Schuch\,\orcidlink{0000-0001-6494-8616}\,}
\affiliation{University of Vienna, Faculty of Physics, Boltzmanngasse 5, 1090 Wien, Austria}
\affiliation{\mbox{University of Vienna, Faculty of Mathematics, Oskar-Morgenstern-Platz 1, 1090 Vienna, Austria}}

\date{\today}

\begin{abstract}

Continuous-time quantum walks on exponentially large, sparse graphs form a
powerful paradigm for quantum computing: On the one hand, they can be
efficiently simulated on a quantum computer. On the other hand, there are \BQP-complete problems based on continuous-time quantum walks, which provides an alternative framework for thinking about quantum
computing---a perspective which has indeed led to a number of novel algorithms
and oracle problems. Recently, simulating the dynamics of a system of harmonic oscillators
(that is, masses and springs) was set forth as another \BQP-complete
problem defined on exponentially large, sparse graphs. In this work, we establish
a mapping between these two classes of problems. As
compared to linking the two classes of problems via their \BQP-completeness, our mapping has several desirable features:
it preserves the structure of the problem, including initialization, read-out, and efficient oracle access, resulting in low overhead in terms of both space and time. It also naturally preserves the geometry of the underlying graph and provides a recipe to translate any quantum algorithm in the quantum walk paradigm to harmonic oscillators (and vice versa), which allows one to map between restricted subsets of instances that are not \BQP-complete, and offers an alternative proof for the \BQP-completeness of the harmonic oscillator problem via a reduction to and from the \BQP-complete quantum walk problem.
\end{abstract}

\maketitle

\section{Introduction}

Quantum walks---that is, the dynamics of a quantum particle (``walker'')
hopping on a graph---have long provided a powerful perspective on quantum
computing, offering alternative ways to devise quantum algorithms~\cite{farhi1998quantum}. A
quantum walk on a graph with $N$ vertices can be encoded using $n=\log_2
N$ qubits, by associating each computational basis state $\ket{v}$ with the
walker being at vertex $v$ of the graph. Based on results from Hamiltonian
simulation, the quantum walk dynamics on an exponentially large graph can
then be simulated efficiently with polynomial resources, provided that the graph is sufficiently sparse and suitably encoded~\cite{berry2007efficient}. 

A key reason for the success of quantum walks as a paradigm for quantum
computing is that they provide a natural approach to a number of
problems, particularly search problems, which can be naturally thought of
as traversing an exponentially wide decision tree. There, the quantum walker
can explore all possibilities simultaneously and---for problems with
suitable structure---determine global properties of such a tree by
interfering signals from all paths. 
In particular, this perspective was 
instrumental in obtaining algorithms for game trees and other problems
which exhibit speed-ups over classical
algorithms~\cite{Farhi_2008,ambainis2014quantumwalkalgorithmelement,magniez2005quantumalgorithmstriangleproblem,buhrman2005quantumverificationmatrixproducts,magniez2007quantumcomplexitytestinggroup,Reichardt_2012},
as well as in
devising a problem (the \emph{glued trees problem}) with provably
exponential oracle separation compared to classical
computers~\cite{Childs_2003}, but it also offered new interpretations of
known algorithms such as Grover search~\cite{Shenvi_2003,childs2004spatial,Szegedy_2004, ambainis2019quadraticspeedupfindingmarked}.  
Conversely, it has also been
shown that any quantum computation can be encoded as a quantum walk
problem on an exponentially large graph: Those quantum walk problems are
\BQP-complete~\cite{Childs_2009}, and thus offer an
alternative, yet equivalent, way of thinking about quantum computations.

In recent work~\cite{Babbush_2023}, a different paradigm of quantum
computing was set forth, namely, simulating the dynamics of an exponential
number of coupled harmonic oscillators (i.e., masses connected by springs),
where the coupling between the masses is once again provided as a sparse,
efficiently encoded graph. The authors proved both that quantum computers
can efficiently simulate exponentially large systems of
masses and springs, and, conversely, that simulating such systems of masses
and springs is \BQP-complete---that is, any quantum algorithm can be
rephrased in this form. In addition, the same glued trees problem as
for quantum walks can be used to prove an exponential oracle separation
for harmonic oscillators. In a closely related vein, it has very recently been established that simulating particle-preserving Gaussian bosonic circuits on exponentially many modes gives rise to \BQP-complete problems \cite{Barthe_2025}, {and that simulating quadratic bosonic Hamiltonians more generally is \BQP-complete, with both quantum walks and harmonic oscillators as special
cases~\cite{zschetzsche2026complexityquadraticbosonichamiltonian},}
further highlighting the connection between universal quantum computation and the dynamics of exponentially large linear systems.

Within this context, there appear to be particularly strong structural
similarities between quantum walks and harmonic oscillator problems as
paradigms for quantum computation. Both problems are defined on
exponentially large, sparse, efficiently encoded graphs. On an intuitive
level, both seem to describe some kind of ``excitation'' moving through this graph: a walker in one case and oscillations---that is, kinetic energy---of a mass in the other
case. Both of these ``excitations''
display interference effects: the quantum walker because of its quantum nature and
the excitation in the mass and spring system because it propagates as a wave.
This raises the question whether there is a deeper structural relation
between the two classes of problems; such a relation might then, in
particular, allow one to intuitively understand the computational power of harmonic
oscillators through quantum walks (or vice versa).

As a first observation, both problems are described by linear differential equations.
Quantum walks are described by first order equations, whereas harmonic oscillators are described by second-order linear differential equations in classical
Newtonian mechanics. Since any second-order equation can be readily transformed to coupled first-order equations by introducing
additional variables---indeed, this corresponds to the
formulation of the problem in the framework of
Hamiltonian mechanics---both problems are described by
a linear differential equation
\begin{equation}
\label{eq:intro-lde}
\dot x(t) = Lx(t)\ ,
\end{equation}
where $x(t)$ is an exponentially large vector, the dot denotes the derivative
with respect to time $t$, and $L$ is a
linear map that characterizes the problem at hand. 
Thus, it is tempting to claim that this proves that the two types of
problems are equivalent, since they are both captured by the same equation
\eqref{eq:intro-lde}.

However, the situation is more subtle than that. Most importantly, note
that the Schrödinger equation for $n$ qubits itself is also nothing but a
first order linear differential equation, Eq.~\eqref{eq:intro-lde}, on a
$2^n$-dimensional vector space, yet proving that simulating either quantum
walks or harmonic oscillator systems is \BQP-complete takes significant
effort.  The challenges are threefold. (\emph{i})~First, each problem
class has its own conditions on the linear operator $L$: In the
Schrödinger equation, $iL$ must be Hermitian,
for a quantum walk, $iL$ must
have non-negative entries,
and for the harmonic oscillator system, $L$ must
effectively encode the second-order Newton equations with positive masses
and spring constants in \eqref{eq:intro-lde}; on top of that, any of these
problems requires suitable, efficient access to $L$, and any mapping must
also map between those access models.  (\emph{ii})~Second, we must be able
to map the initial states onto another in a way that stays within the
class of allowed initial states.  (\emph{iii})~Third, we must be able to
efficiently map the output of one problem class to the other, whether it
is a sample drawn from some distribution, a numerical value, or the answer
to a decision problem.  

In this paper, we resolve these challenges by establishing an equivalence that does not rely on intermediate circuit-model simulations. We define our mapping as having three core properties:

\emph{Direct}: We transform the exponential-sized differential equation of the quantum walk into the Hamiltonian equation of the harmonic oscillators (and vice versa) via explicit algebraic substitutions, making sure the resulting operators $L$, initial conditions, and read-outs respect the conditions outlined in (\emph{i})-(\emph{iii}).

\emph{Transparent}: The reduction is constructed via explicit, localized steps that preserve the underlying structure of the problems (see Table \ref{table}). Because the graph is only modified locally, the scaling behavior of each component remains tightly bounded: 
the graph size, degree, and number of non-zero initial components scale predictably as low-order linear or polynomial functions of the original network parameters, and the output distributions are related through an explicit transition matrix with at most a polynomial number of non-zero terms. Crucially, this localized structure ensures our mappings remain strictly compatible with black-box oracle access to the sparse graphs.

\emph{Physically intuitive}: The spatial geometry of the underlying graph is preserved (requiring only the addition of specific nearest and next-nearest neighbor edges), allowing the quantum interference of the walker to be physically understood via the classical wave propagation of the springs (and vice versa).

While the existence of \BQP-complete problems based on both quantum walks and harmonic oscillators already implies the existence of polynomial-time reductions between them, such generic reductions do not preserve the natural structure of the problem. Specifically, the existing proof for the \BQP-completeness of the harmonic oscillator problem~\cite{Babbush_2023} relies on a Kitaev clock construction and requires spectral analysis to prove that the desired outcome can be extracted by repeated sampling across multiple time steps. {Furthermore, their construction does not establish any correspondence to a continuous-time quantum walk. In contrast to this approach,} our mapping between quantum walks and harmonic oscillators bypasses the intermediate circuit model entirely by establishing a direct algebraic equivalence between the dynamics of the two problems. This yields a proof of \BQP-hardness of the harmonic oscillator problem by reduction from Childs’ \BQP-hardness construction for quantum walks~\cite{Childs_2009} that explicitly preserves the structure of the underlying graph.
Our direct mapping has the additional advantage that it applies beyond \BQP-complete variants: It allows one to map any specific instance of a quantum walk problem to an instance of a harmonic oscillator problem, and vice versa, so it natively provides mappings for restricted subclasses (and potentially more general classes) of the two problems.

Let us briefly provide some details about the mappings between the graphs, summarized in Fig.~\ref{fig:flowchart}.
Both mappings consist of two steps. In the first step, the construction
maps the graph of a quantum walk to a graph of a harmonic oscillator
system, where both the nearest and the next nearest neighbors of the original graph
are connected. Conversely, the graph of a harmonic oscillator system
is mapped to a graph of a quantum walk, where each edge is decorated with an
additional vertex. The resulting graphs have the right structure for the
corresponding problem, but do not yet exhibit the correct sign pattern
required by the respective problem. This is corrected for in the second
step, which we term the sign-split embedding: We embed the graph in
a graph of twice the size, constructed from two copies of the original
graph, where edges with negative weights now become edges between the two copies;
given an initial state with opposite amplitudes in the two copies, this
exactly mimics the dynamics of the original graph with negative edge
weights, using only positive weights.

The paper is structured as follows: In Section~\ref{sec:definitions}, we
introduce the necessary concepts, including the model used 
to query exponentially large, sparse graphs (Sec.~\ref{sec:graphs-oracles}), and the different ways
to characterize the output of the problems considered (Sec.~\ref{sec:problems-outputs-variants}), and we define quantum 
walk problems (Sec.~\ref{sec:qw}) and harmonic oscillator problems
(Sec.~\ref{sec:definition-ho}). In Section~\ref{sec:reductions}, we
provide the sought-after mappings between quantum walks and harmonic
oscillator systems. After a summary of the problem setting, we explain
the sign-split embedding used in both mappings (Sec.~\ref{sec:sign-trick}),
and subsequently show how to map quantum walks to harmonic oscillators
(Sec.~\ref{sec:qw<ho}) and vice versa (Sec.~\ref{sec:ho<qw}). Finally, we
conclude in Section~\ref{sec:conclusion}.

\begin{figure*}
    \centering
    \begin{tikzpicture}[
        box/.style={draw, rounded corners=2pt, minimum height=1cm, align=center,
                    inner sep=6pt, font=\small},
        qwbox/.style={box, fill=Blue!15, draw=Blue!60!black},
        hobox/.style={box, fill=Orange!15, draw=Orange!70!black},
        arr/.style={-{Stealth[length=2.5mm]}, thick},
        rowlabel/.style={font=\bfseries\small}
    ]
        \node at (3, 1.7) {\color{Blue!60!black}\bfseries Quantum Walk};
        \node at (11.5, 1.7) {\color{Orange!70!black}\bfseries Harmonic Oscillators};

        \node[rowlabel] at (-1.5,0.5) {(a)};
        \node[qwbox, minimum width=3cm] (T) at (1.5,0.5) {Adjacency Matrix $T$};
        \node[qwbox, minimum width=3cm] (Tt) at (5.5,0.5) {$\tilde T = T + \gamma \mathds{1}$};
        \node[hobox, minimum width=3cm] (At) at (9.5,0.5) {$\tilde A = \tilde T^2$};
        \node[hobox, minimum width=3cm] (A) at (13.5,0.5)
            {$A = \begin{pmatrix} A_d & - A_o \\ - A_o & A_d \end{pmatrix}$};

        \draw[arr] (T) -- (Tt);
        \draw[arr] (Tt) -- (At);
        \draw[arr] (At) -- (A);

        \node[rowlabel] at (-1.5,-2) {(b)};
        \node[qwbox, minimum width=3cm, minimum height=1.6cm] (T2) at (1.5,-2)
            {$T = \begin{pmatrix}
                0&0&B_p&B_n \\
                0&0&B_n&B_p \\
                B_p^\dagger & B_n^\dagger&0&0 \\
                B_n^\dagger & B_p^\dagger&0&0
            \end{pmatrix}$};
        \node[hobox, minimum width=3cm] (B) at (5.5,-2) {$B = B_p - B_n$};
        \node[hobox, minimum width=3cm] (AB) at (9.5,-2) {$A = B^\dagger B$};
        \node[hobox, minimum width=3cm] (A2) at (13.5,-2) {Coupling Matrix $A$};

        \draw[arr] (B) -- (T2);
        \draw[arr] (AB) -- (B);
        \draw[arr] (A2) -- (AB);
    \end{tikzpicture}
    \caption{\textbf{Flowchart summarizing the construction of the dynamical generator in each reduction direction.} Row (a), quantum walk $\to$ harmonic oscillator: starting from the adjacency matrix $T$, a diagonal shift $\tilde T = T+\gamma\mathds{1}$ with $\gamma=3d$ ensures positivity of $\tilde T$, after which the coupling matrix of the oscillator system is built from $\tilde A=\tilde T^2$ after being split into its diagonal and off-diagonal parts, $A_d$ and $A_o$, and recombined via the sign-split embedding. Row (b), harmonic oscillator $\to$ quantum walk: the coupling matrix $A$ is factorized as $A=B^\dagger B$ via an incidence matrix $B\in\Mat(dN \times N, \mathds{R})$, which is split into its entry-wise positive and negative parts $B=B_p-B_n$, and recombined via the sign-split embedding into the adjacency matrix $T$ of the quantum walk.}
    \label{fig:flowchart}
\end{figure*}

\section{Problem definitions and setup}

\label{sec:definitions}
In the following, we will start by introducing both problems: quantum
walks and harmonic oscillators. Crucially, we will see that both
problems can be described by real first-order linear differential
equations $\dot x(t) = L x(t)$, where $x(t)\in\mathbb R^N$ is a vector in
a real vector space, the dot denotes the derivative with respect to $t$,
and $L$ is a real-valued matrix. The difference between the two classes of
problems lies in the specific properties required of the matrix $L$, as
well as in the specific choice of initial conditions and final readout. 

The fact that these are first-order linear differential equations means that they can be simulated efficiently using quantum mechanical systems with $n\sim\log N$ qubits~\cite{Berry2014}.
$N$ is therefore exponential in the system
size $n$, so an efficient encoding of $L$ as well as the initial conditions
and the readout is required, where ``efficient'' refers to a
polynomial scaling in $n$ rather than $N$. In both cases, this is
achieved by constructing $L$ from an exponentially large graph that is
efficiently encoded, which is where we start our discussion.

\subsection{Graphs and oracle access\label{sec:graphs-oracles}}

\subsubsection{Graphs}

The problems we consider are constructed from exponentially large weighted graphs $G=(V,E)$. A
weighted graph consists of a set of vertices $v\in V$, a set of edges
$e\in E$, with $e=(v,w)$ connecting vertices $v$ and $w$, and weights
$t(e)\in\mathbb R$ associated with each $e\in E$. Here, we also allow for
self-loops $(v,v)\in E$. The matrix $T_{vw}=t((v,w))$ with $v,w\in V$,
$T_{vw}=T_{wv}$, is
called the \emph{(weighted) adjacency matrix} of the graph. The degree
$d$ of the graph is the largest number of vertices to which any
single vertex is connected, $d = \max_v \big|\{w\in V:(v,w)\in E\}\big|$, where $|\cdot|$ denotes the size of a set.  We define $N=|V|$, and $n=\log N$.

\subsubsection{Efficient encoding and oracle access\label{sec:oracle-access}}

Since we will be concerned with exponentially large graphs, we will require that
the adjacency matrix $T$ (and thus the graph) can be efficiently encoded.
Specifically, we require that the degree $d=\poly(n)$---that is, each row of $T$ has at most $\poly(n)$ non-zero entries---and that there is an efficient way to query the non-zero entries in each row. 

We formalize this through an \emph{oracle}, i.e., a function treated as a
black-box, which takes as input a vertex $v$ and an integer $i \in \{1, ..., d \}$ and returns the column $w$ and the value $T_{vw}$ of the $i$-th non-zero entry in row $v$ (which can be specified to exponential precision in $T_{vw}$ using $\poly(n)$ bits). Note that access to such an oracle allows one to
determine all the vertices adjacent to any given vertex $v$ using $\mathcal{O}(d)$ oracle queries.

In a concrete problem setting, this oracle will be specified by an efficient
classical circuit of size $\poly(n)$, that is, a concrete function that can be
computed in $\poly(n)$ time.  In a more abstract setting, we can treat it as a
black-box function whose computational cost we don't consider. Rather, what we
care about is how often we have to query the oracle to solve the problem at
hand or to map it to another one. 

\subsubsection{Functions of oracles}

In order to map between different graph-based problems, we will need to transform oracles.
Given oracle access to a set of matrices $\{T^i_{vw}\}_i$ (of size
exponential in $n$), we want to provide oracle access to some matrix-valued
function $f(T^1,\dots,T^k)=M$. This oracle access must be efficient, that is, one query to the oracle for $M$ must only require $\poly(n)$ queries to the oracles for $T^i$. This guarantees overall efficiency if the oracle itself is given by an efficiently computable function. Assuming oracle access to both the rows and 
columns of the matrices $T^i$ (which is automatic for Hermitian matrices), we demand row and column access to $M$.

Some functions that allow efficient oracle access are:

(i) Addition: $M=T^1+T^2$. 

(ii) Multiplication: $M=T^1T^2$. Given $v$, the
non-zero $M_{vw}=\sum_u T^1_{vu}T^2_{uw}$ can be obtained by first
querying the row
oracle to determine the set $U^1$ of all $u$ with $T^1_{vu}\ne0$ and their
value. Then, the non-zero elements of $M_{vw}=\sum_{u \in U^1} T^1_{vu}T^2_{uw}$ can be computed with $|U^1|=\poly(n)$ queries to the row oracle for $T^2$.

(iii) Matrices in block form, where we have efficient oracle access to each block.

(iv) Entry-wise functions $M_{vw}=f(T_{vw})$ for any efficiently
computable $f$ with $f(0)=0$, such as picking only positive or
negative entries, or the entry-wise square root. 

\subsection{Outputs and their relation: Sampling, estimating probabilities, and
decision problems\label{sec:problems-outputs-variants}}

\subsubsection{Problem variants}

For the problems we are considering, the output is characterized by a
probability distribution $P(s)$ over a discrete set of outcomes $s\in
S$. In the context of running problems on a quantum computer, $P(s)$ corresponds to the probability distribution obtained when measuring the
output register in the computational basis.
Depending on the required output format, these problems naturally manifest in two distinct complexity-theoretic frameworks:

\begin{enumerate}
\item 
\emph{The sampling variant:} Solving the problem involves sampling from the
output distribution $P$, i.e., returning an outcome $s$ drawn according to
$P(s)$. Samples can be drawn either from the exact distribution
or from one that is $\varepsilon = 1/\poly(n)$ close; note that these
cannot be distinguished in $t=\poly(n)$ time if $\varepsilon$ can be chosen to scale as {$1/\poly(t)$}. This can be formally captured through the complexity class $\mathsf{SampBQP}$ (Sampling \BQP) \cite{Aaronson2013}.
\item 
\emph{The estimation variant:} The simulation outputs an estimate of
$P(S_0)=\sum_{s\in S_0}P(s)$, for an efficiently characterizable subset
$S_0\subset S$ of outcomes (i.e., where one can decide efficiently
whether $s\in S_0$) specified as part of the input, up to precision
$1/\poly(n)$. This can be formally captured through the complexity class $\mathsf{FBQP}$ (Functional \BQP) \cite{Aaronson2013}.
\item[2a.]
\emph{The decision version:} Based on the estimation variant, we can also construct a decision version of the problem (which belongs to \BQP \ and can be used to define \BQP-complete problems).
The ``yes'' and ``no'' instances correspond to the cases $P(S_0)\ge
a$ and $P(S_0)\le b$, respectively, for any given $a$, $b$, with $a-b\ge 1/\poly(n)$. 
\end{enumerate}

Importantly, these variants are equivalent via Cook reductions: given access to a black box
that solves one variant, we can construct an efficient black box that solves
the other variant (which runs in $\poly(n)$ time and queries
the black box $\poly(n)$ times). 

Sampling can be used to estimate probabilities up to $1/\poly(n)$
precision with an exponentially small failure
probability. Conversely, estimating
probabilities of subsets $S_0$ can be used to sample from
$P(s)$ up to $1/\poly(n)$ precision, e.g., by thinking
of $s$ as a bit string and sampling bit by bit by choosing suitable
subsets $S_0$.  

Clearly, access to an estimate of $P(S_0)$ with
precision $a-b=1/\poly(n)$ allows one to solve the decision variant with
thresholds $a$ and $b$ (with an exponentially small failure probability
inherited from the estimation variant). Conversely, the decision version can be used to obtain a $1/\poly(n)$ estimate of $P(S_0)$ by binary search over $a$ and $b$.

\subsubsection{Mapping between different problems}

The goal of this work is to relate two different computational problems,
quantum walk problems and harmonic oscillator problems, in either variant. Since both variants are fully characterized by the underlying probability distribution $P(s)$, we
will take a high-level approach and construct reductions that
explicitly relate the output distributions. Those relations
readily allow one to obtain mappings between the different variants of the two
problems.

Let the two problems be described by probability distributions $P(s)$,
$s\in S$ and $Q(r)$, $r\in R$. To show that the problem for
$P(s)$ can be simulated through the problem for $Q(r)$, we
construct an efficient mapping that yields a
relation of the form
\begin{equation}
\label{eq:prob-dist-relation}
P(s) = \sum_{r\in R} C_{sr} Q(r)\ .
\end{equation}
The matrix $C_{sr}$ that relates these two output distributions must satisfy
$\sum_s C_{sr}=1$ for all $r\in R$, so that probabilities are conserved. Because $S$ and $R$ are exponentially large in $n$, we 
require (and show) that $C_{sr}$ is both row- and column-sparse, that is, each
row and column has at most $\poly(n)$ non-zero entries, and that both rows and
columns of $C$ can be queried efficiently (cf.~Section~\ref{sec:oracle-access}).
Importantly, $C_{sr}$ can have negative entries, as we will see in one of the reductions. This is rooted in the fact that the output distribution $Q(r)$ of the 
harmonic oscillator problem is highly over-parameterized and thus highly constrained, such that $\sum_r C_{sr}Q(r)\ge0$ still holds for all allowed output distributions $Q(r)$.

A relation of the form \eqref{eq:prob-dist-relation} between the output
distributions of two problems establishes a very strong connection between them,
as it allows one to simulate the problem described
by $P(s)$ through access to the problem described by $Q(r)$.

If $C_{sr}\ge0$, it yields an immediate mapping between the sampling
variants: given a sample $r\in R$ drawn from $Q(r)$, the algorithm
outputs $s$ with probability $C_{sr}$; then, $s$ is distributed
according to $P(s)$. This constitutes a many-one reduction
between sampling problems: a single query to the sampling oracle for
$Q$, followed by a fixed randomized post-processing step, analogous to
a randomized Karp (many-one) reduction for decision problems.

If $C_{sr}\not\ge0$, access to an oracle that returns estimates of
$Q(r)$ with $1/\poly(n)$ precision still allows one to
estimate $P(s)$ by querying the oracle to determine the set
$R_r=\{r\in R|C_{sr}\ne0\}$, and subsequently computing
$P(s) = \sum_{r\in R_r} C_{sr} Q(r)$ (where the accuracy of the $Q(r)$
is chosen accordingly), as long as the $|C_{sr}|$ are bounded by
$\poly(n)$. In that case, the resulting reduction between the
estimation problems (which are functional problems) is analogous to a
truth-table reduction for decision problems (that is, a Cook reduction
with non-adaptive queries, where the oracle is queried a polynomial
number of times in parallel).

Finally, Eq.~\eqref{eq:prob-dist-relation} allows us to simulate the
estimation variant for $P(s)$ through the sampling variant of $Q(r)$,
even if $C_{sr}\not\ge0$: Given sampling access to $Q(r)$ and a subset
$S_0\subset S$, we can for each sample $r_i$, $i=1,\dots,I$, query the
column oracle for $C_{sr_i}$ to determine the positions $s$ of its
non-zero entries, and estimate $P(S_0)$ through
{$\tfrac1I \sum_i \sum_{s \in S_0} C_{s r_i}$ up to $1/\poly(n)$} precision with
$I=\poly(n)$ samples, as long as $|C_{sr}|$ is bounded by $\poly(n)$. In
combination with the equivalence between the different variants for
each problem discussed above, this implies that any variant of one
problem can be simulated through the other even if $C_{sr}\not\ge0$.

As a corollary, since decision versions can be mapped to and from the
estimation variant, this also yields an efficient mapping between the
decision versions of the problem. If direct access to the
estimation oracle for $Q$ is available, this reduction is non-adaptive:
an estimate of $P(S_0)$ is obtained using $\poly(n)$ non-adaptive
queries as described above, which is then compared against the
thresholds $a,b$. If only the decision oracle for $Q$ is available,
each of the $\poly(n)$ relevant values of $Q(r)$ can instead be
estimated to $1/\poly(n)$ precision using $O(\log n)$ adaptive queries
via binary search over the thresholds of $Q$'s decision problem, which
can in turn be simulated using $O(\poly(n))$ non-adaptive queries. This
results in a total of $O(\poly(n))$ non-adaptive queries; for graphs of
bounded or $O(\log n)$ degree, where at most $O(\log n)$ values of $Q(r)$
are relevant, this can be reduced further to $O(\mathrm{polylog}(n))$
adaptive queries. The resulting reduction is thus a truth-table
reduction, that is, a Cook reduction (where the oracle can be queried
$O(\poly(n))$ times) restricted to non-adaptive queries; for graphs of
bounded degree, the reduction can be further restricted to a Cook
reduction with $O(\mathrm{polylog}(n))$ adaptive queries.

\begin{table*}
\begin{center}
\renewcommand{\arraystretch}{1.4}
\begin{tabular*}{\linewidth}{@{\extracolsep{\fill}} l c c @{\extracolsep{\fill}}}
    \toprule \toprule
     & \textbf{Quantum Walk} & \textbf{Harmonic Oscillator} \\
    \midrule \midrule
    Graph & order $2^n$, degree $d=\poly(n)$ & order $2^n$, degree $d=\poly(n)$ \\
    Characterized by &
        \makecell{real, symmetric adjacency matrix \\ $T_{vw}=T_{wv}\geq 0$} &
        \makecell{spring constants $\kappa_{vw}=\kappa_{wv}\geq 0$ \\ Laplacian $A_{vw}=-\kappa_{vw}$, $A_{vv}=\sum_w \kappa_{vw}$} \\
    \midrule
    Amplitude encoding & vertex population $\abs{a_v(t) + i b_v(t)}^2$ & displacements $q(t)$, momenta $p(t)$ \\[0.3em]
    Evolution equations &
    $\begin{pmatrix} \dot{a} \\ \dot{b} \end{pmatrix}
    = \begin{pmatrix} 0 & T \\ -T & 0 \end{pmatrix}
    \begin{pmatrix} a \\ b \end{pmatrix}$
    &
    $\begin{pmatrix} \dot{q} \\ \dot{p} \end{pmatrix}
    = \begin{pmatrix} 0 & \mathds{1} \\ -A & 0 \end{pmatrix}
    \begin{pmatrix} q \\ p \end{pmatrix}$ \\[0.6em]
    Initial conditions  & $k=\poly(n)$ non-zero, real amplitudes & $q(0)=0$, $k=\poly(n)$ non-zero momenta $p_i(0)$ \\
    \midrule
    Output distribution &
    $P_{\mathrm{QW}}(v) = \big|\langle v\ket{\Phi(t_\mathrm{final})}\big|^2 = |c_v|^2$, $v\in V$
    &
    {\footnotesize $ P_{\mathrm{HO}}(s) = \dfrac{1}{\mathcal H} \left\{ \begin{array}{l@{\;}l}
	\tfrac12 p_v^2\ , & s=v\in V\\[0.6ex]
	\tfrac12 \kappa_{vw} (q_v-q_w)^2\ ,& s=(v,w)\in E \\[0.6ex]
	\tfrac12 \kappa_{vv} q_v^2\ ,& s=(v,v)\in E
\end{array}\right.$} \\
    \bottomrule
\end{tabular*}
\end{center}
\caption{Structural correspondence between the quantum walk and harmonic oscillator formalisms, showing matched graph parameters, evolution equations, initial conditions, and output distributions. Here $c_v = a_v + ib_v$ denotes the complex amplitude at vertex $v$, $\mathcal{H} = \frac{1}{2}p^Tp + \frac{1}{2}q^TAq$ is the conserved total energy, and $k$ denotes the number of non-zero entries in the initial state. All variables are defined in detail in Section~II.}
\label{table}
\end{table*}

\subsection{Quantum walks}
\label{sec:qw}

\subsubsection{Setup and evolution equation}

A (continuous-time) quantum  walk describes the
dynamics of a single quantum particle hopping on a weighted graph $G=(V,E)$.  The
state of the system is thus described by a state
\begin{equation}
\label{eq:rwalk-complex-wavefunction}
\ket{\Phi(t)} = \sum_{v\in V} c_v(t) \ket{v}\in \mathbb C^{|V|}\ ,
\end{equation}
where the orthonormal basis vectors $\ket v$ correspond to the particle
being located at vertex $v$, and $\sum_v \vert c_v \vert^2 = 1$. The dynamics of the system is then given
by the weighted adjacency matrix $T$ of the graph through the Schrödinger
equation
    $\ket{\dot{\Phi}(t)} = -iT \ket{\Phi(t)}$,
or, in terms of $c=(c_v)_{v\in V}$, 
\begin{equation} 
    \dot{c} = -iTc\ ,
    \label{eq:pde-randwalk-complex-c}
\end{equation}
which is a first order linear differential equation in a complex vector space.
(Note that we will frequently omit the explicit time dependence $c_v(t)$ and
simply write $c_v$ in the following, as long as it is clear from the context.) An adaptation for the use of the Laplacian can be found in Appendix~\ref{app:laplacian}.

Importantly, for a quantum walk, we require $T$ to be symmetric with
non-negative  entries, $T_{vw}=T_{wv}$, $T_{vw}\ge0$. As the underlying graph is
generally exponentially large in $n$, $|V|=N$ with $n=\log N$, we
demand that $T$ be efficiently encoded, that is, the graph has degree 
$d=\poly(n)$ and $T$ can be efficiently queried, as discussed above.

Equation \eqref{eq:pde-randwalk-complex-c} can be re-expressed as a
differential equation over a real vector space by parameterizing $c_v = a_v
+ i b_v$. Inserting this into \eqref{eq:pde-randwalk-complex-c} gives rise
to the equation
\begin{align*}
\dot a_v + i\, \dot b_v  &= 
-i\sum_w T_{vw} (a_w  +i\, b_w )
\\
&= \sum_{w} (-i T_{vw} a_w  + T_{vw} b_w) \ .
\end{align*}
Since the entries of $T$ are real, this can be decoupled into two
sets of real differential equations by separating the real and imaginary part.
We then arrive at the first-order real differential equation 
\begin{align}
    \begin{pmatrix}
        \dot{a} \\
        \dot{b}
    \end{pmatrix}
    =
    \begin{pmatrix}
        0 & T \\
        -T & 0
    \end{pmatrix}
    \begin{pmatrix}
        a \\
        b
    \end{pmatrix}\ .
    \label{eq:pde-randwalk-real}
\end{align} 
Importantly, since $T$ is a real symmetric matrix, the generator of the
evolution in the equation above is antisymmetric, and the resulting
evolution will thus preserve the norm $\sum_v (a_v^2+b_v^2)$.

For completeness, let us note that by taking the time derivative and re-substituting the l.h.s.\ of \eqref{eq:pde-randwalk-real} on the right, we can arrive at 
\emph{decoupled} second-order linear differential equations for $a$ and $b$,
$\ddot a = -T^2 a$ and $\ddot b=-T^2 b$, which are related by the
condition $\dot b= -Ta$. We could therefore also take these second-order
equations as our working definition (which suggest a connection to the 
Newton equations for harmonic oscillators).

\subsubsection{Initialization}

One natural initial state for the quantum walk is $\ket{\Phi(0)}=\ket{v}$, where the particle is placed at one specific vertex $v\in V$ initially. 
More broadly, if the quantum walk is simulated on a quantum computer, any state $\sum_v c_v\ket{v}$ that can be efficiently prepared by a quantum circuit constitutes a meaningful initial state. This includes superpositions of small numbers of vertices, $\ket{\Phi(0)}=\sum_{v\in \mathcal I} c_v\ket{v}$, where the
set $\mathcal I$ is at most of size $\poly(\log|V|)$. On a quantum computer, where the basis
$\{\ket{v}\}_{v\in V}$ is encoded in  $n=\log|V|$ qubits, this is a
superposition of $\poly(n)$ basis states and can therefore be prepared
efficiently~\cite{Gleinig_2021}.
In what follows, we restrict ourselves to initial states with purely real amplitudes.

\subsubsection{Output}

The output of the quantum walk problem corresponds to a measurement of the final
evolved state $\ket{\Phi(t_\mathrm{final})} = \sum_v c_v\ket{v}$ in the vertex
basis $\{\ket{v}\}_{v\in V}$. The set of all outcomes is thus equal to the
vertex labels $V$ of the underlying graph, with probability distribution
\begin{equation}
P_{\mathrm{QW}}(v) = \big|\langle v\ket{\Phi(t_\mathrm{final})}\big|^2
    = |c_v|^2 \ ,\quad v\in V\ .
\label{eq:qwalk-output-distribution}
\end{equation}

\subsubsection{Formal problem statement: Quantum Walk Problems, and BQP-complete variants}

Let us now summarize the formal problem statement for the class of Quantum
Walk Problems.

A \emph{Quantum Walk Problem} is specified by the adjacency matrix $T$ with
\begin{equation}
    T_{vw}=T_{wv}, \ T_{vw}\ge 0 
    \label{eq:conditions-on-T}
\end{equation} 
of a weighted graph $G=(V,E)$ on $N$
vertices with maximum degree $\poly(n)$, $n=\log N$, where we have oracle
access to $T$, together with an initial state $\ket{\Phi(0)}\in \mathbb
C^{|V|}$. The output of the problem is then characterized by the
probability distribution \eqref{eq:qwalk-output-distribution} over the set
of vertices $V$, where
$\ket{\Phi(t_\mathrm{final})}=e^{-it_\mathrm{final}T}\ket{\Phi(0)}$ is
obtained by integrating Eq.~\eqref{eq:pde-randwalk-complex-c} for a time
$t_{\mathrm{final}}$.  

For suitable choices of the initial state and $T$, the decision version of
the problem (cf.\ Section~\ref{sec:problems-outputs-variants}) is
\BQP-complete.  Specifically, the Quantum Walk Problem with an efficient
(i.e., $\poly(n)$ time) circuit for querying $T$, $T_{vw}=\poly(n)$,
initial state $\ket{\Phi(0)}=\sum_{v\in \mathcal I} c_v(0)\ket{v}$ with
$|\mathcal I|=\poly(n)$ and $c_v(0)\in\mathbb R$, and
$t_\mathrm{final}=\poly(n)$ is \BQP-complete. The problem remains
\BQP-complete if the initial state is restricted to a
single vertex $v\in V$.

This \BQP-completeness follows immediately from Childs' work on quantum
computing by quantum walks~\cite{Childs_2009}, where $T$ is the
adjacency matrix of an unweighted graph of constant maximum degree that can
be queried efficiently. The output
distribution is obtained by measuring in the vertex basis at $t_\mathrm{final}=\poly(n)$, and accepting (rejecting) if $P(S_0) > a$ {($P(S_0) < b$)}, where $S_0$ is a succinctly
characterized set of vertices (the ``output wire'' corresponding to the
``yes'' output of the encoded circuit), and $a$, $b$ both scale like
$1/\poly(n)$.

Note that we can always obtain $0\le T_{vw}\le 1$ by rescaling
$t_{\mathrm{final}}\rightsquigarrow t_{\mathrm{final}}\times\max_{v,w}T_{vw}$.

\subsection{Harmonic oscillators\label{sec:definition-ho}}

\subsubsection{Setup and evolution equation}

Classical harmonic oscillators describe systems of pointlike particles (``masses'') connected by springs to one another and to some static wall (i.e., points fixed in space). The system can be defined in terms of a weighted graph $G=(V,E)$, where edge weights $\kappa_{vw}=\kappa_{wv}>0$ with $v\neq w$ represent spring constants between masses $v$ and $w$, and self-loops with weights $\kappa_{vv} > 0$ represent the couplings of a mass $v$ to the wall. Absent edges indicate the absence of a spring, or $\kappa_{vw}=0$. For simplicity, we assume unit masses ($m_v=1$ for all $v$); the general case only requires a straightforward rescaling of $\kappa_{vw}$ that preserves the evolution equations and is compatible with oracle access, cf.~Ref.~\cite{Babbush_2023}.

In the framework of Hamiltonian mechanics, the state of the system is
fully characterized by vectors of conjugate variables: the position or displacement $q(t)=(q_v(t))_v\in
\mathbb R^{|V|}$ and the momentum $p(t)=(p_v(t))_v$.
The Hamiltonian function
\begin{equation}
\label{eq:hosc-hamiltonian-kappa}
\mathcal{H}=\frac{1}{2}\sum_v p_v^2+\frac{1}{2}
\sum_{(v,w)\in E}\!\!\kappa_{vw} (q_v-q_w)^2 + \frac12\sum_v \kappa_{vv} q_v^2\
\end{equation}
specifies the total energy of the system (the analog of
$\bra{\Phi(t)}H\ket{\Phi(t)}$), where in the middle term, each edge only
adds one term to the sum.
By defining a matrix $A$ with $A_{vw} = - \kappa_{vw}$ for $v\ne w$ 
and $A_{vv}=\sum_w \kappa_{vw}$, 
we can express \eqref{eq:hosc-hamiltonian-kappa} as
\begin{equation}
    \mathcal{H}=\tfrac{1}{2}\,p^Tp+\tfrac{1}{2}\,q^TAq\ .
    \label{eq:hosc-hamiltonian-A}
\end{equation}
Its evolution is governed by the Hamilton
equations $\dot{q}=\frac{\partial\mathcal{H}}{\partial p}$ and 
$\dot{p}=-\frac{\partial\mathcal{H}}{\partial q}$, where the derivatives with respect to vectors $p$ and $q$ are gradients.

This gives rise to the following real first order differential equation
describing the dynamics of the system: 
\begin{align}
    \begin{pmatrix}
        \dot{q} \\
        \dot{p}
    \end{pmatrix}
    =
    \begin{pmatrix}
        0 & \mathds{1} \\
        -A & 0
    \end{pmatrix}
    \begin{pmatrix}
        q \\
        p
    \end{pmatrix}\ .
        \label{eq:hosc-pde}
\end{align}
Importantly, in order to represent a physical spring system, where spring constants $\kappa_{vw}$ are non-negative, $A$ must be a
symmetric matrix with non-positive off-diagonal entries,
$A_{vw}=A_{wv}=-\kappa_{vw}\le0$ for $v\ne w$, and $\sum_{w}A_{vw}=\kappa_{vv}\ge 0$. 
Unlike the evolution equation of the quantum walk,
Eq.~\eqref{eq:hosc-pde} is not norm-preserving; rather it preserves the
total energy $\mathcal H$ given by Eq.~\eqref{eq:hosc-hamiltonian-A}, since Eq.~\eqref{eq:hosc-pde} implies $\dot{\mathcal H}=0$.

Again, let us note for completeness that Eq.~\eqref{eq:hosc-pde} can be converted to a
second-order differential equation for $q$ by substituting the l.h.s.\ on
the right, giving rise to $\ddot q = -Aq$. This is precisely the equation
obtained in classical mechanics by combining Newton's equation $m_v\ddot
q_v = F_v$, where $F_v$ is the force acting on mass $v$, with Hooke's law $F_v =
-\kappa_{vv} q_v + \sum_{w \ne v} \kappa_{vw} (q_w-q_v)$.

Let us briefly point out an interesting aspect of the conditions imposed
on $A$.  On the one hand, these imply that $A$ is positive semi-definite,
which is equivalent to the system being stable (i.e., $\ddot q=-Aq$ having
oscillating rather than exponentially growing solutions, or equivalently
the energy function $\mathcal H$ being lower semi-bounded, which is a
property satisfied by any physical system of masses and springs). However,
positive semi-definiteness of $A$ does \emph{not} imply that $A$ describes
a system of springs with only non-negative spring constant---somewhat
surprisingly, there can be systems, where some ''springs'' have negative
spring constant, that are nevertheless stable, and we provide some
illustrative examples in appendix~\ref{app:sign-separation}.

Finally, let us remark that the equations above can also be used to model systems
in higher spatial dimensions (e.g., systems of masses and springs in
three-dimensional space), where each mass is described by several
coordinates $q_v$ and $p_v$: Since the different spatial coordinates are
not coupled by the evolution equation, these decouple into separate equations for each spatial coordinate. Note that for a real isotropic system, the
mass and spring constant must be the same for all spatial directions,
which imposes additional constraints on $A$.

\subsubsection{Initialization}

The initialization of the harmonic oscillator system involves fixing the
initial values of position and momentum, $\big( {q(0)\atop p(0)}\big)$.
Again, we are primarily interested in situations where such an initial
condition can be efficiently specified and prepared.  Typically, these are
initial states for which a $\poly(n)$ number of $q_i(0)$ and $p_i(0)$ are
non-zero. However, when running the problem on a quantum computer, any
initial state that can be prepared by a $\poly(n)$-size quantum circuit
in the encoding used will suffice.

\subsubsection{Output}

At the end of the simulation, we must specify a set of outcomes with an associated probability distribution. For a harmonic oscillator system, these outcomes correspond to the degrees of freedom in which the system's energy is stored: (1) the masses, which hold kinetic energy, and (2) the springs (either between the masses or to the walls), which hold potential energy. These correspond to the first versus the second and third groups of terms in Eq.~\eqref{eq:hosc-hamiltonian-kappa}, respectively.

The set of outcomes $S$ of the read-out procedure is thus given by the vertices and edges of the underlying graph $G=(V,E)$, such that $S=V\cup E$. The probability $P_{\mathrm{HO}}(s)$ for each outcome $s\in S$ is defined to be proportional to its stored energy:
\begin{equation}
\label{eq:output-prob-HO}
P_{\mathrm{HO}}(s) = \frac{1}{\mathcal H}\left\{ \begin{array}{l@{\quad}l} 
	\tfrac12 p_v^2\ , & s=v\in V\\[1ex]
	\tfrac12 \kappa_{vw} (q_v-q_w)^2\ ,& s=(v,w)\in E \\[1ex]
	\tfrac12 \kappa_{vv} q_v^2\ ,& s=(v,v)\in E
\end{array}\right.\ .
\end{equation}
Because the total energy $\mathcal{H}$ of the system, Eq.~\eqref{eq:hosc-hamiltonian-A}, is conserved, this mapping naturally satisfies the required conservation of probability.

\subsubsection{Formal problem statement: Harmonic Oscillator Problems and
BQP-complete variants}

We now summarize the formal problem statement for the class of Harmonic Oscillator Problems.

A \emph{Harmonic Oscillator Problem} is specified by the adjacency matrix
$\kappa_{vw}=\kappa_{wv}$, $\kappa_{vw}\ge0$ of a weighted graph $G=(V,E)$
with self-loops on $N$ vertices with maximum degree $d=\poly(n)$, $n=\log(N)$,
where we have oracle access to $\kappa$, together with an initial state
$\big({q(0)\atop p(0)}\big)$. Define $A_{vw}:=-\kappa_{vw}$ for $v\ne w$, $A_{vv}=\sum_w \kappa_{vw}$. The output of the problem is then characterized by
the probability distribution \eqref{eq:output-prob-HO} over the set $S=V\cup E$
of vertices and edges (i.e., masses and springs), where $q$ and $p$ in
\eqref{eq:output-prob-HO} are obtained by integrating Eq.~\eqref{eq:hosc-pde}
for a time $t_\mathrm{final}$.

Note that there is a 1-to-1 correspondence between symmetric
$\kappa_{vw}\ge0$ and $A_{vw}$ with
\begin{equation}
A_{vw}=A_{wv}\,,\ A_{vw}\le0\mbox{\ for\ }v\ne w\,,\ \sum_w A_{vw}\ge
0\ .
\label{eq:conditions-on-A}
\end{equation}
Furthermore, an efficient oracle for $A$ can be used to construct an
efficient oracle for $\kappa$ and vice versa. We can therefore equally use $A$
as the defining weighted adjacency matrix for the Harmonic Oscillator
Problem.

For suitable choices of the initial state and $\kappa$, the decision
version of the problem (cf.\ Section~\ref{sec:problems-outputs-variants}) is
\BQP-complete. Specifically, the 
Harmonic Oscillator Problem with an efficient (i.e., $\poly(n)$ time)
circuit for querying $\kappa$, $\kappa_{vw}=\poly(n)$, and an initial state
$\big({0\atop p(0)}\big)$, with $p_i(0)=0$ except for $\poly(n)$ values of $i$,
and $t_\mathrm{final}=\poly(n)$, is \BQP-complete. Furthermore, it 
remains \BQP-complete if the initial state is restricted to two non-zero entries
$p_i(0)$.

The \BQP-completeness of both problems follows immediately from
Theorem~1 and Theorem~3 in Ref.~\cite{Babbush_2023}.

\section{Reductions: Mapping between quantum walks and harmonic oscillators}
\label{sec:reductions}

Our goal is to show that the Quantum Walk Problem and the Harmonic
Oscillator Problem can mutually simulate each other. To achieve this, we
will establish direct mappings between the differential equations
that describe the two problems. For convenience, we recall these equations below:
For the Quantum Walk Problem, we have the complex differential equation~\eqref{eq:pde-randwalk-complex-c}
\begin{equation*}
    \dot{c} = -iTc\ ,
    \tag{\ref{eq:pde-randwalk-complex-c}}
\end{equation*}
or, equivalently, the real differential equation \eqref{eq:pde-randwalk-real}
\begin{equation*} 
    \begin{pmatrix}
        \dot{a} \\
        \dot{b}
    \end{pmatrix}
    =
    \begin{pmatrix}
        0 & T \\
        -T & 0
    \end{pmatrix}
    \begin{pmatrix}
        a \\
        b
    \end{pmatrix}\ ,
    \tag{\ref{eq:pde-randwalk-real}}
\end{equation*}
where $T$ satisfies the constraints specified in Eq.~\eqref{eq:conditions-on-T}. 
For the Harmonic Oscillator Problem, we have~\eqref{eq:hosc-pde},
\begin{equation*}
    \begin{pmatrix}
        \dot{q} \\
        \dot{p}
    \end{pmatrix}
    =
    \begin{pmatrix}
        0 & \mathds{1} \\
        -A & 0
    \end{pmatrix}
    \begin{pmatrix}
        q \\
        p
    \end{pmatrix}\ ,
    \tag{\ref{eq:hosc-pde}}
\end{equation*}
where $A$ satisfies the constraints specified in Eq.~\eqref{eq:conditions-on-A}.

Although both problems are described by linear differential equations, the
structure of the linear operator differs, and a key challenge will be to find a mapping that respects the conditions imposed on $T$ and $A$. Additionally, we will need to understand how these
maps transform initial conditions and read-outs, and
whether these transformations can be performed efficiently, in particular for
the \BQP-complete versions of the problems.

On a formal level, it is suggestive to choose $A=T^2$ and perform a
corresponding change of variables. While efficient oracle access to the inverse mapping $T=\sqrt{A}$ is impossible in general, a workaround is to define $T$ via a rectangular matrix $B$ where $A=B^\dagger B$~\cite{Babbush_2023}.
However, this $T$ and $A$ fail the required constraints. Furthermore, it is unclear how the
change of variables affects the initial conditions and read-out, particularly because it may require the inverse of $T$.
This section demonstrates how to overcome these challenges.

\subsection{The sign-split embedding\label{sec:sign-trick}}

The two matrices $T$ and $A$, which appear in the differential equations
describing the two problems, require specific sign patterns, which
will not be immediately present after the transformation. To
recover the correct sign pattern, we will make use of the following
trick, which we term the \emph{sign-split embedding}. 

First, let us introduce some formalism. Consider a vector space $\mathcal
V\equiv \mathbb C^{|S|}$ containing vectors $x=(x_s)_{s\in S}\in \mathcal
V$, with some (finite) index set $S$.  We then introduce a space
$\bar{\mathcal V}$ of twice the dimension, which we view as two copies of
$\mathcal V$, $\bar{\mathcal V} = \mathcal V \oplus \mathcal V$.  A vector
$z\in\bar{\mathcal V}$ is of the form
\begin{equation}
\label{eq:double-space-z-equals-x-y}
z = \begin{pmatrix} x \\ y \end{pmatrix}\ ,
\end{equation}
with $x,y\in \mathcal V$. Correspondingly, the
index set $\bar S$ of $\bar{\mathcal V}$ contains two copies of the
index set $S$. We
denote the maps that embed the index set $S$ into the first
and second copy in $\bar S$ by $\sigma_1:S\to\bar S$ and
$\sigma_2:S\to\bar S$, respectively, and the inverse map by $\pi:\bar{S}\to S$. In
Eq.~\eqref{eq:double-space-z-equals-x-y}, for $s\in S$ we have $z_{\sigma_1(s)}=x_s$,
$z_{\sigma_2(s)}=y_s$, and $\pi(\sigma_1(s))=\pi(\sigma_2(s))=s$.

Now consider the linear differential equation 
\begin{equation}
\dot x(t) = M\, x(t)\ ,
\label{eq:signtrick-original-pde}
\end{equation}
where $x(t)\in \mathcal V$.
Let $M=P-N$. Then, in the doubled space $\bar{\mathcal V}$,
\begin{equation}
\begin{pmatrix}P & N \\ N & P \end{pmatrix}
\begin{pmatrix} \phantom{-}x \\ -x \end{pmatrix}
= 
\begin{pmatrix} Px-Nx \\ Nx-Px \end{pmatrix}
= 
\begin{pmatrix} \phantom-Mx \\ -Mx \end{pmatrix}\ ,
\label{eq:signtrick-preserve}
\end{equation}
that is, $\big(\begin{smallmatrix} P&N\\N&P\end{smallmatrix}\big)$ leaves
the space of antisymmetric vectors invariant, and acts as $M$ on each
of the two components.

Now consider the differential equation
\begin{equation}
 \dot z(t) 
= \begin{pmatrix}P & N \\ N & P \end{pmatrix}
z(t)\ ,
\label{eq:signtrick-doubled-pde} 
\end{equation}
where $z(t)\in \bar{\mathcal V}$.  Furthermore, let us choose an 
antisymmetric initial condition $z(0)=\big({x(0)\atop-x(0)}\big)$.
Then, \eqref{eq:signtrick-preserve} implies that the evolution of the
system will stay within this antisymmetric space,
$z(t)=\big({x(t)\atop-x(t)}\big)$, or
$$
\begin{pmatrix} \phantom-\dot x(t) \\ -\dot{x}(t) \end{pmatrix}
= 
\begin{pmatrix}P & N \\ N & P \end{pmatrix} 
\begin{pmatrix} \phantom-x(t) \\ -x(t) \end{pmatrix}\ ,
$$ 
where $x(t)$ is nothing but the solution of the original differential equation
\eqref{eq:signtrick-original-pde} with initial condition $x(0)$. 

We have thus succeeded in obtaining the solution of the differential equation
\eqref{eq:signtrick-original-pde} from the solution of the doubled
differential equation \eqref{eq:signtrick-doubled-pde}, simply by choosing
antisymmetric initial conditions. 
This doubling technique structurally parallels the bipartite double cover standard in algebraic graph theory \cite{BrouwerHaemers2012}. {A related construction, in which negative matrix entries are handled by an ancilla prepared in $\ket{-}$ \cite{Jordan_2010, Childs_2015}, is also used in the \BQP-hardness proof of
Ref.~\cite{Babbush_2023}.} We adapt this concept here to change the sign structure of the linear operator in the
differential equation: For example, if $M$ is real, decomposing it into non-negative parts $P$ and $N$ with $M=P-N$ ensures that the new operator in Eq.~\eqref{eq:signtrick-doubled-pde} contains only
non-negative entries. The trick can be easily generalized to also
transform complex operators into such a form by using four copies.

Let us note that the sign-split embedding still applies if $M$ has a
block-structure, where applying the trick to $M$ as a whole is equivalent
to applying it to each block individually (as we will do in the following).

\subsection{Simulating quantum walks by harmonic oscillators}
\label{sec:qw<ho}

In this section, we provide a reduction of the Quantum Walk Problem to
the Harmonic Oscillator Problem, that is, we show how the Quantum Walk
Problem can be simulated through a Harmonic Oscillator Problem. To this
end, we consider the Quantum Walk Problem on a graph $G=(V,E)$ with
adjacency matrix $T$.

\subsubsection{Mapping of dynamics\label{sec:qw-to-ho-dynamics}}

We start by showing how we map the dynamics of the Quantum Walk
Problem---that is, the differential equation
\eqref{eq:pde-randwalk-complex-c}, or equivalently
\eqref{eq:pde-randwalk-real}---to the Harmonic Oscillator
Problem~\eqref{eq:hosc-pde}. Following our definition of the Quantum Walk
Problem, we will require $T$ to have entries $0\le T_{vw}\le 1$, but the
construction we give readily generalizes to $T$ with entries $-1\le T_{vw}\le 1$.

The construction of the mapping proceeds in several steps.  First, observe
that replacing $T$ by $\tilde T:=T+\gamma\openone$ in
Eq.~\eqref{eq:pde-randwalk-complex-c} introduces a constant energy shift
$\gamma$ and thus a global phase $e^{-i\gamma t}$ to the complex
amplitudes $c_v$ in the solution of the complex Schrödinger equation
\eqref{eq:pde-randwalk-complex-c}. It thus describes a fully  equivalent
problem (except in an unphysical setting in which the read-out is sensitive
to the global phase).

Specifically, we will choose $\gamma=3d$ in the following (with
$d=\poly(n)$ the maximum degree of the graph), a choice that will serve several
purposes. For now, since the spectral radius of $T$ is bounded by $d$, it ensures that $\tilde T$ is invertible.

Now consider \eqref{eq:pde-randwalk-real} with
$T$ replaced by $\tilde T:=T+\gamma\openone$. 
Since $\tilde T$ is invertible, we can introduce new
variables $\left({q\atop p}\right)$ with
\begin{equation}
\begin{pmatrix}a\\b\end{pmatrix} = 
\begin{pmatrix}\tilde T q \\ p\end{pmatrix}\ .
    \label{eq:rwalk-to-hosc-substitution}
\end{equation}
Substituting this in the real differential equation \eqref{eq:pde-randwalk-real}
for the quantum walk results in the differential equation
$$
 \begin{pmatrix} \dot{q} \\ \dot{p} \end{pmatrix}
    = \begin{pmatrix} 0 & \mathds{1} \\ -\tilde T^2 & 0 \end{pmatrix}
    \begin{pmatrix} q \\ p \end{pmatrix}\ ,
$$
which is already of the desired form \eqref{eq:hosc-pde}, except that
$\tilde A:=\tilde T^2$ does not satisfy the required conditions
\eqref{eq:conditions-on-A}.  Quite on the contrary, 
as all entries of $\tilde T$ are non-negative, the same will hold for
$\tilde A$, and thus condition \eqref{eq:conditions-on-A} will necessarily be
violated. 

To fix this issue, we use the sign-split embedding.  Let $A_d$ and $A_o$
be the diagonal and off-diagonal part of $\tilde A=A_d+A_o$, respectively,
and define\footnote{In case $-1\le T_{ij}\le 1$, $A_o$ only contains 
the positive off-diagonal entries, and the subsequent arguments need to be adapted accordingly.}
\begin{equation}
    A:=
    \begin{pmatrix}
        A_d & -A_o\\
        -A_o & A_d
    \end{pmatrix}\ .
    \label{eq:A_of_QW}
\end{equation} 
This $A$ satisfies the conditions \eqref{eq:conditions-on-A}: $A_{vw}\le0$
for $v\ne w$ holds trivially because $\tilde T$ is entry-wise non-negative.
That $\sum_w A_{vw}\ge0$ holds follows from $\gamma=3d$,
by combining the following observations: The
column sum of $A_o$ is upper bounded by the column sum of $T^2+2\gamma T$; the
column sum of $T^2$ corresponds to the sum of all distance-two
paths from that vertex, so it is upper bounded by $d^2$;
the column sum of $T$ is upper bounded by $d$; 
and finally, the entries of $A_d$ are lower bounded by
$\gamma^2$.
(Note that adding $\gamma$ is indeed required 
since otherwise, e.g., \ for a fully connected (sub-)graph where $T_{ij}=1\
\forall\,i,j$, $\sum_w A_{vw}<0$.) In fact, with our choice of $\gamma$,
we have $\kappa_{vv}=\sum_w A_{vw}\ge 2d^2$; in particular,
$\kappa_{vv}$ is bounded below by a constant for any graph. At the same
time, the diagonal of $\tilde T^2$ is bounded above by $\gamma^2+d^2$, and
thus $\kappa_{vv}=\sum_w A_{vw} \le A_{vv} \le \gamma^2+d^2=\poly(n)$; we will
make use of both of these bounds later on. 

The sign-split embedding comes with a doubling of the index set $V$ of
vertices to $\bar{V}$. By choosing antisymmetric initial conditions, the state of the doubled system is at all times of the form
\begin{equation}
    \begin{pmatrix} \bar{q}\\ \\ \bar{p} \end{pmatrix}
=
    \begin{pmatrix}
	\phantom-q \\ -q \\ \phantom-p \\ -p
    \end{pmatrix}\ ,
\label{eq:rwalk-to-hosc-antisymstate}
\end{equation} 
and its dynamics is governed by the differential equation
\begin{equation}
\label{eq:qw-to-ho-full-ho-pde}
    \begin{pmatrix}
	\phantom-\dot{q} \\ -\dot{q} \\ 
	\phantom-\dot{p} \\ -\dot{p}
    \end{pmatrix}=
    \begin{pmatrix}
        0 &0 &\mathds{1} &0\\
        0 &0 &0 &\mathds{1}\\
        A_d & -A_o &0 &0\\
        -A_o & A_d &0 &0
    \end{pmatrix}
    \begin{pmatrix}
	\phantom-q \\ -q \\ \phantom-p \\ -p
    \end{pmatrix}
\ ,
\end{equation} 
which is just Eq.~\eqref{eq:hosc-pde}, with $A$ from
\eqref{eq:A_of_QW}, and thus describes a Harmonic Oscillator Problem.  The
solution of this equation will thus---under the substitution
\eqref{eq:rwalk-to-hosc-substitution}---be precisely that of the quantum
walk problem~\eqref{eq:pde-randwalk-real} (up to the global phase rotation
due to $\gamma$). Eq.~\eqref{eq:qw-to-ho-full-ho-pde} also implicitly
defines the edges $\bar E$ of the underlying graph $\bar G = (\bar V,\bar E)$
of the Harmonic Oscillator Problem.

Note that the total energy of this harmonic oscillator system---relevant
for the normalization of the output probabilities---equals
$$
\bar{\mathcal H}  = \tfrac12 \bar p^T\bar p + \tfrac12 \bar q^TA\bar q
= p^Tp + q^T \tilde A q
= b^Tb + a^Ta = 1\ ,
$$
using that $\left(a \atop b\right)$ describes the normalized initial state
of a quantum walk.

Finally, following the discussion in Section~\ref{sec:graphs-oracles}, note that 
oracle access to $A$, Eq.~\eqref{eq:A_of_QW}, can be efficiently
constructed given oracle access to $T$, where one query to the oracle for $A$ requires $\mathcal{O}(d^2)$ queries to the oracle for $T$. In particular, this implies that,
for Quantum Walk Problems where an efficient circuit for $T$ is provided,
an efficient circuit for $A$ in the Harmonic Oscillator Problem can be
built as well.

\subsubsection{Initial state preparation}

Now, let us show how and under what conditions we can map an initial
state $c(0)=a(0)+ib(0)$ of the quantum walk to the corresponding initial
condition $\big({{\bar q(0)}\atop \bar p(0)}\big)$ of the harmonic oscillator system. 
The general relation is obtained by combining the substitution
\eqref{eq:rwalk-to-hosc-substitution} with the antisymmetric encoding
\eqref{eq:rwalk-to-hosc-antisymstate}, which results in
the relation
\begin{equation}
\label{eq:qp-from-ab-general}
\begin{pmatrix} \bar q\\[2.5ex] \bar p \end{pmatrix}
=
    \begin{pmatrix}
	\phantom-q \\ -q \\ \phantom-p \\ -p
    \end{pmatrix}
= 
    \begin{pmatrix}
	\phantom-\tilde{T}^{-1} a \\ -\tilde{T}^{-1} a \\ \phantom- b \\ - b
    \end{pmatrix}
\end{equation}
between the initial conditions. The difficulty now lies in the fact
that, in general, there is no efficient way to apply $\tilde T^{-1}$,
and suitable restrictions are needed to solve 
\eqref{eq:qp-from-ab-general} for
$\big({\bar q\atop \bar p}\big)$, given $c=a+ib$.

To this end, we exploit the fact that $c=(c_v)_v$ is fixed only up to a global
phase $c_v\leftrightarrow e^{i\phi}c_v$. Thus, if all $c_v(0)$ point in the same direction in the complex plane, we can use the global phase
to make $c$ purely imaginary, i.e., $c(0)=i\,b(0)$, and $a(0)=0$. This is in
particular the case for our definition of the Quantum Walk Problem, where the initial conditions are real, $c_v(0)\in\mathbb R$. In that case, we can choose the initial
condition of the harmonic oscillator system as
\begin{equation}
\label{eq:qp-initial-conditions}
\begin{pmatrix} \bar q(0)\\[2.5ex] \bar p(0) \end{pmatrix}
= 
    \begin{pmatrix}
	0 \\ 0 \\ c(0) \\ -c(0)
    \end{pmatrix}\ .
\end{equation}
If $c(0)$ has a small (e.g.,
$\mathrm{poly}(n)$) number of non-zero entries or can otherwise be efficiently prepared on a quantum computer, then the same is true for the initial state $\big({\bar q(0)\atop \bar p(0)}\big)$ of the Harmonic
Oscillator Problem.

This implies that any real-valued initial condition $c(0)$ of a Quantum Walk
Problem with $k$ non-zero entries maps to an
initial state $\big({\bar q(0)\atop \bar p(0)}\big)$ of the corresponding Harmonic
Oscillator Problem with $2k$ non-zero entries, and similarly for initial
conditions which can be efficiently prepared by a quantum circuit. In
particular, this means that the initial conditions of the \BQP-complete version
of the Quantum Walk Problem are mapped to initial conditions covered by the
\BQP-complete version of the Harmonic Oscillator Problem.

\subsubsection{Read-out}

In the Quantum Walk Problem, the read-out is a measurement in the
computational basis with a probability distribution
$P_{\mathrm{QW}}(v) = |c_v|^2$. 
We will now show how to relate this to the output distribution $P_{\mathrm{HO}}(s)$ of the Harmonic Oscillator Problem using a mapping that is transparent and efficient both in the setting in which we want to sample from $P_{\mathrm{QW}}$ and in the setting in which we want to estimate $P_{\mathrm{QW}}(v^\ast)$ for a given vertex $v^{\ast}$.

As discussed in Section~\ref{sec:definition-ho}, the set $S$ of possible
outcomes of the Harmonic Oscillator Problem consists of all edges
(springs) and all vertices (masses), $S=\bar E\cup \bar V$, with the probability
$P_\mathrm{HO}(s)$ of outcome $s\in S$
given by Eq.~\eqref{eq:output-prob-HO}. We can solve
\eqref{eq:output-prob-HO} to obtain
\begin{subequations}
\label{eq:hosc-output-qq-pp-qmixed}
\begin{equation}
\label{eq:hosc-output-qq-pp}
\bar p_{\bar v}^2 =  
2P_{\mathrm{HO}}(\bar v)\ ,
\quad
\bar q_{\bar v}^2 =  
\frac{2P_{\mathrm{HO}}((\bar v,\bar v))}{\kappa_{\bar v\bar v}}\ ,
\end{equation}
for $\bar v \in \bar V$ and $(\bar v,\bar v)\in \bar E$, respectively, and
\begin{equation}
\label{eq:hosc-output-qvqw}
\bar q_{\bar v}\bar q_{\bar w} = 
\frac{P_{\mathrm{HO}}((\bar v,\bar v))}{\kappa_{\bar v\bar v}}
+\frac{P_{\mathrm{HO}}((\bar w,\bar w))}{\kappa_{\bar w\bar w}}
-\frac{P_{\mathrm{HO}}((\bar v,\bar w))}{\kappa_{\bar v\bar w}}
\end{equation}
\end{subequations}
for $(\bar v,\bar w)\in \bar E$,  $\bar v \ne \bar w$. Both $\kappa_{\bar v\bar
w}=-A_{\bar v\bar w}$ and $\kappa_{\bar v\bar v}=\sum_w A_{\bar v\bar w}$ can be
efficiently obtained given oracle access to $T$.
We then have that for $v\in V$,
\begin{equation}
\label{eq:pQW-from-qq-pp}
\begin{aligned}
P_{\mathrm{QW}}(v) & = |c_v|^2 = a_v^2 + b_v^2 
\\ 
& \stackrel{\eqref{eq:rwalk-to-hosc-substitution}}{=} 
    ((\tilde Tq)_v)^2 + p_v^2
\\ &= 
    \sum_{w,u} \tilde T_{vw} \tilde T_{vu} q_w q_u + p_v^2
\\
& \stackrel{\eqref{eq:rwalk-to-hosc-antisymstate}}{=}
    -\sum_{w\ne u} \tilde T_{vw}\tilde T_{vu} \bar q_{\sigma_1(w)} \bar q_{\sigma_2(u)}
\\
& \qquad + \frac{1}{2} \sum_{w} \tilde T_{vw}^2 \left( \bar q_{\sigma_1(w)}^2 + \bar q_{\sigma_2(w)}^2 \right) \\
& \qquad + \frac{1}{2} \left( \bar p_{\sigma_1(v)}^2 + \bar p_{\sigma_2(v)}^2 \right) \ .
\end{aligned}
\end{equation}
In the last step, we have used that 
$q_w=\bar q_{\sigma_1(w)}=-\bar q_{\sigma_2(w)}$,
Eq.~\eqref{eq:rwalk-to-hosc-antisymstate}, to pick pairs $\bar q_{\bar
w}\bar q_{\bar u}$ for which $(\bar w,\bar u)\in \bar E$, and which 
can thus be obtained using
Eqs.~(\ref{eq:hosc-output-qq-pp},\ref{eq:hosc-output-qvqw}).
Note that, since 
$\bar q_{\sigma_1(w)}=-\bar q_{\sigma_2(w)}$ and $\bar p_{\sigma_1(w)}=-\bar
p_{\sigma_2(w)}$, there are multiple ways of writing
Eq.~\eqref{eq:pQW-from-qq-pp}, and we choose it symmetric in the two copies.

By combining Eq.~\eqref{eq:pQW-from-qq-pp} with
Eq.~\eqref{eq:hosc-output-qq-pp-qmixed}, we find that
$P_{\mathrm{QW}}(v) = \sum_{s} C_{vs} P_\mathrm{HO}(s)$, where oracle access to
the columns as well as the rows of the transition matrix $C_{vs}$ can be
efficiently implemented given oracle access to the adjacency matrix $T$
characterizing the quantum walk. As we show in appendix~\ref{app:distributions},
$C_{vs}$ satisfies $\sum_v C_{vs}=1$, but it can possess negative entries (which
we illustrate in the appendix with a concrete example). The latter relates to
the fact that the output distribution $P_\mathrm{HO}$ of the Harmonic Oscillator Problems is
highly over-parametrized and thus highly constrained, guaranteeing
positive output probabilities $P_\mathrm{QW}$ despite a non-positive transition matrix
$C_{vs}$. ($P_\mathrm{HO}$ depends only on the $2|\bar{V}|$ independent positions and
momenta, but can have up to $\bar d|\bar V|$ edges, with $\bar d$ the degree of
the underlying graph.)  Yet, as we also discuss in
appendix \ref{app:distributions}, the entries of $C_{vs}$ are nevertheless bounded, and therefore, one
can still use oracle access to $C_{vs}$ to map from a $1/\poly(n)$ estimate of
$P_\mathrm{HO}$ to a $1/\poly(n)$ estimate of $P_\mathrm{QW}$.

\begin{center}
   \textasteriskcentered\quad\textasteriskcentered\quad\textasteriskcentered
\end{center}

Putting all pieces together, we can apply the discussion of
Section~\ref{sec:problems-outputs-variants}, which shows that solving the
Quantum Walk Problem can be reduced to solving the Harmonic Oscillator Problem.
In particular, this yields a direct reduction of the \BQP-complete version of
the Quantum Walk Problem to the \BQP-complete version of the Harmonic Oscillator
Problem, where, given an initial state with $k$ non-zero entries of $c(0)$ in
the former problem, an initial state with $2k$ non-zero entries of $p(0)$ is
required in the latter. We summarize the results of the reduction in the following theorem:
\begin{theorem}
    Any Quantum Walk Problem on an efficiently encoded graph with $N=2^n$ vertices and maximum degree $d=\poly(n)$, with an initial state of $k=\poly(n)$ non-zero real amplitudes, can be efficiently simulated by a Harmonic Oscillator Problem with $2N$ vertices (masses) and a maximum degree of $\mathcal{O}(d^2)$, with an initial state $\left({q(0)\atop p(0)}\right)$ with $q(0)=0$ and $2k$ non-zero momenta $p_i(0)$. Simulating a single query to the oracle for $A$ requires $\mathcal{O}(d^2)$ queries to the oracle for $T$.
    To map the decision problems, $\poly(n)$ queries to the decision oracle for the Harmonic Oscillator Problem and $\mathcal{O}(d^3)$ queries to the oracle for $T$ are required to solve the decision version of the Quantum Walk Problem.
\end{theorem}

\subsection{Simulating harmonic oscillators by quantum walks}
\label{sec:ho<qw}
In this section, we provide a reduction of the Harmonic Oscillator
Problem to the Quantum Walk Problem, that is, we show how the Harmonic
Oscillator Problem can be simulated by mapping it to a Quantum Walk
Problem.

\subsubsection{Mapping of dynamics}

We start by constructing a mapping from the dynamics of the Harmonic
Oscillator Problem, Eq.~\eqref{eq:hosc-pde}, to the complex differential
equation of the Quantum Walk Problem,
Eq.~\eqref{eq:pde-randwalk-complex-c}. A key challenge is that, while the natural
mapping from $A$ to $T$ is $T=\sqrt{A}$, the square root is
incompatible with oracle access to $A$, even given an efficient implementation of
that oracle. To resolve this issue, we use the results of Ref.~\cite{Babbush_2023}, notably the decomposition of $A$ into $A=B^\dagger B$ with $B$ an incidence matrix of the graph, to rephrase the harmonic oscillator equation as a Schrödinger equation with a real-valued Hamiltonian. 
As an additional step we embed the resulting differential equation into a larger Hilbert space using the sign-split embedding to obtain a quantum walk equation of the form \eqref{eq:pde-randwalk-complex-c}, where $T$ satisfies the constraints specified in Eq.~\eqref{eq:conditions-on-T}. Through a suitable choice of initial conditions, we can ensure
that the dynamics in this larger space remain confined to a suitable
subspace in which there is a one-to-one correspondence between the
dynamics of the quantum walk and the harmonic oscillators.

The construction of the matrix $B$ and the mapping to a Schrödinger equation can be found in Ref.~\cite{Babbush_2023} and, for completeness, is repeated in appendix \ref{sec:oscillators_to_se} using notation that is in line with the rest of this work.
The resulting differential equation is the following: 
\begin{equation}
\begin{pmatrix} \dot c^E(t) \\ \dot c^V(t) \end{pmatrix}
=
-i
\underbrace{\begin{pmatrix} 0 & B \\ B^\dagger & 0
\end{pmatrix}}_{=:\tilde T}
\begin{pmatrix} c^E(t) \\ c^V(t) \end{pmatrix}\ ,
\tag{\ref{eq:hosc-to-qwalk-schroedingereq}}
\end{equation}
with
\begin{equation}
\begin{pmatrix} c^E(t)\\c^V(t)\end{pmatrix}
=
\frac{1}{\sqrt{2\mathcal H}}
\begin{pmatrix} -iB q(t) \\ p(t) \end{pmatrix}\ .
\tag{\ref{eq:hosc-to-qwalk-substitution}}
\end{equation}
We report here some essential facts about this construction; the derivation is given in Appendix~\ref{sec:oscillators_to_se}. The
quantities $c^E$ and $c^V$ are complex amplitudes associated with the
edges and vertices of the original graph $G=(V,E)$, respectively. As
seen in Eq.~\eqref{eq:hosc-to-qwalk-schroedingereq}, the matrix $B$
couples $c^E$ and $c^V$; here, $B_{ev}$ is non-zero only when vertex $v$
is incident to edge $e$, so that $B$ is sparse, and oracle access to $B$
can be efficiently constructed from oracle access to $\kappa$. Finally, if the initial condition is chosen such that $c^V(0)$ is real and
$c^E(0)$ lies in $i\,\mathrm{Im}_{\mathbb R}B$ (in particular, purely
imaginary), this property is preserved under the time
evolution~\eqref{eq:hosc-to-qwalk-substitution}.

In order to transform Eq.~\eqref{eq:hosc-to-qwalk-schroedingereq} to
a Quantum Walk Problem, we still need to ensure that all entries in the linear
operator are positive, while $\tilde T$, by construction of $B$ \eqref{eq:def-B}, also has 
negative entries. This is once more achieved by using the sign-split  embedding: By splitting $B=B_p-B_n$ into a positive and negative part,
doubling the degrees of freedom, and restricting to antisymmetric initial
conditions, we arrive at a quantum walk
\begin{equation}
\begin{pmatrix} \phantom-\tfrac{1}{\sqrt{2}}\dot{c}^E \\ -\tfrac{1}{\sqrt{2}}\dot{c}^{E} \\ 
\phantom-\tfrac{1}{\sqrt{2}}\dot{c}^V  \\ -\tfrac{1}{\sqrt{2}}\dot{c}^{V} \end{pmatrix}
=
-i
\underbrace{\begin{pmatrix} 
0 & 0 & B_p & B_n  \\
0 & 0 & B_n & B_p  \\
B_p^\dagger & B_n^\dagger & 0 & 0 \\
B_n^\dagger & B_p^\dagger & 0 & 0 
\end{pmatrix}}_{=:T}
\begin{pmatrix} \phantom-\tfrac{1}{\sqrt{2}}c^E \\ -\tfrac{1}{\sqrt{2}}c^{E} \\ 
\phantom-\tfrac{1}{\sqrt{2}}c^V  \\ -\tfrac{1}{\sqrt{2}}c^{V} \end{pmatrix}\:,
\label{eq:hosc-to-qwalk-quantumwalk}
\end{equation}
whose dynamics are equivalent to \eqref{eq:hosc-to-qwalk-schroedingereq},
and which is of the desired form \eqref{eq:pde-randwalk-complex-c} of the 
Quantum Walk Problem.  (Here, the $\tfrac{1}{\sqrt{2}}$ ensures normalized
vectors.) Note that efficient oracle access to $T$ can be constructed
given efficient oracle access to $B$, where one query to the oracle for $T$ requires a single query to the oracle for $\kappa$, or $\mathcal{O}(d)$ queries to the oracle for $A$.

\subsubsection{Initial state preparation}
Following Eq.~\eqref{eq:hosc-to-qwalk-substitution} combined with the
doubling of the system due to the sign-split embedding, any initial
condition of the Harmonic Oscillator Problem with $k$ non-zero entries
can be transformed to an initial condition of the Quantum Walk Problem
with at most $2dk$ non-zero entries, where $d$ is the degree of the graph.
In particular, the mapping between such encodings can be implemented
efficiently given oracle access to $\kappa$ (and thus $B$). For
$k=\poly(n)$ non-zero entries, this gives rise to an initial state for the
Quantum Walk Problem with $\poly(n)$ non-zero entries. If the $k$
non-zero entries of the Harmonic Oscillator Problems are all in $p(t)$,
then this results in an initial condition for the Quantum Walk Problem
with $2k$ non-zero entries, which are all purely real.

In particular, this maps initial conditions between the \BQP-complete versions of the Harmonic Oscillator Problem and the Quantum Walk Problem.

\subsubsection{Read-out}

For the read-out, the key observation is that with the relation
\eqref{eq:hosc-to-qwalk-substitution}, we have that for an edge
$e=(w,u)\in E$,
\begin{equation}
\big|c^E_e\big|^2 
= 
\frac{1}{\mathcal H}\left\{ \begin{array}{l@{\quad}l} 
	\tfrac12 \kappa_{wu} (q_w-q_u)^2\ ,& w \neq u \\[1ex]
	\tfrac12 \kappa_{ww} q_w^2\ ,& w=u
\end{array}\right.\ ,
\end{equation}
while for $v\in V$,
\begin{equation}
\big|c^V_v\big|^2 
= 
\frac{1}{2\mathcal H}\, p_v^2\ .
\end{equation}
We thus find that the amplitudes squared of the individual components of
$c$ are in one-to-one correspondence to the energy stored in the
individual springs and masses of the harmonic oscillator system,
which in turn precisely give the output probability
distribution of the Harmonic Oscillator Problem. 

To express  the output distribution $P_{\mathrm{HO}}(s)$ with $s\in S=E\cup
V$ in terms of $P_{\mathrm{QW}}$, we just need to note that the Quantum
Walk Problem in Eq.~\eqref{eq:hosc-to-qwalk-quantumwalk} lives in a
doubled Hilbert space $\bar S$ due to the sign-split embedding. The relation
between the distributions is thus given by 
\begin{equation}
P_{\mathrm{HO}}(s) = \tfrac12(P_\mathrm{QW}(\sigma_1(s)) +
P_\mathrm{QW}(\sigma_2(s)))\ ,
\end{equation}
which has the required form~\eqref{eq:prob-dist-relation}, with a stochastic
transition matrix $C_{sr}\ge0$. 

\begin{center}
   \textasteriskcentered\quad\textasteriskcentered\quad\textasteriskcentered
\end{center}

Together with  the discussion in Section~\ref{sec:problems-outputs-variants},
this proves that the task of solving the Harmonic Oscillator Problem can be
reduced to solving the Quantum Walk Problem. In particular, it provides a direct
reduction of the \BQP-complete version of the Harmonic Oscillator Problem to
the \BQP-complete version of the Quantum Walk Problem. Specifically, given a
Harmonic Oscillator Problem with $k$ non-zero entries of the initial state
$\big({q(0)\atop p(0)}\big)$, this requires an initial state with $2dk$ non-zero
entries $c(0)$ for the corresponding Quantum Walk Problem, with $d$ the degree
of the underlying graph; if for the initial state of the Harmonic Oscillator
Problem, $q(0)=0$, this reduces to $2k$ non-zero real-valued coefficients
$c(0)$. We summarize the results of the reduction in the following theorem:
\begin{theorem}
    Any Harmonic Oscillator Problem on an efficiently encoded graph with $N=2^n$ vertices (masses), $\vert E \vert$ edges (springs), and maximum degree $d=\poly(n)$, with an initial state $\left({q(0)\atop p(0)}\right)$ with $q(0)=0$ and $k=\poly(n)$ non-zero momenta $p_i(0)$, can be efficiently simulated by a Quantum Walk Problem with $2(N+\vert E \vert)$ vertices and a maximum degree of $d$, where the initial quantum walk state has $2k$ non-zero amplitudes, all of which are real. Simulating a single query to the oracle for $T$ requires a single query to the oracle for $\kappa$, or $\mathcal{O}(d)$ queries to the oracle for $A$.
    To map the decision problems, $\poly(n)$ queries to the decision oracle for the Quantum Walk Problem are required to solve the decision version of the Harmonic Oscillator Problem.
\end{theorem}

\section{Conclusion\label{sec:conclusion}}
In this paper, we have considered two natural problems for quantum computers defined on exponentially large, efficiently encoded, sparse graphs: The
Quantum Walk Problem, which involves simulating continuous-time quantum walks on a graph, 
and the Harmonic Oscillator Problem, which involves simulating the dynamics of classical
harmonic oscillator systems with harmonic couplings (springs) given by a
graph, and have constructed direct mappings between these two classes of problems: our mapping transforms a Quantum Walk Problem on a given graph
into a Harmonic Oscillator Problem on a graph obtained by connecting the original graph's nearest and next-nearest neighbors with
suitable weights. The reverse mapping converts the graph of the Harmonic Oscillator
Problem into a graph for the Quantum Walk Problem by inserting an additional
vertex on each edge (= spring). In both cases, the resulting graphs have
both positive and negative edge weights, which is subsequently resolved by the
sign-split embedding: A doubling of the graph, with positive-weight edges
connecting vertices in the same (in different) copies if their weight prior to
doubling was positive (negative).

While for both classes of problems, \BQP-complete variants have been identified,
which in particular implies the equivalence of those specific variants of the two
problems, our mapping has a number of advantages: 
First, by using localized steps, the mapping is highly transparent and preserves the evolution equations, initial conditions, and read-outs in a way that is compatible with efficient oracle access.
Second, it exposes the structural similarities between the two paradigms, offering a physically intuitive mapping between the underlying graphs that natively preserves their geometry.
Third, because the correspondence is direct, relying on explicit algebraic substitutions rather than intermediate circuit models, it establishes an equivalence between individual instances and restricted subclasses of either problem which are not \BQP-complete. This direct nature allows one to transform any problem of interest defined in
one setting to the other setting. For instance, any problem known to provide a
quantum speed-up or an oracle separation in the quantum walk setting readily
yields a corresponding problem for harmonic oscillators, beyond the glued-trees
construction of Ref.~\cite{Babbush_2023}; e.g., NAND-tree evaluation can be
mapped explicitly onto a network of coupled classical oscillators.

The mapping we devise is not limited to physical systems of masses and springs
but also allows for negative spring constants as long as the overall system is
stable, i.e., its energy is bounded from below (which amounts to a unitary
evolution). It is thus also applicable to harmonic systems which arise from
expanding a general classical system around its equilibrium position to second
order, and thus encompasses a diverse range of scenarios in classical physics.

\begin{acknowledgments}
The authors thank Alice Barthe 
for helpful discussions.  N.S.\ would like to thank the Simons
Institute for the Theory of Computing for their hospitality during the
Summer Cluster on Quantum Computing in July 2023, where the 
motivation to study this question originated.
This work has been funded in parts by the Austrian Science Fund FWF (Grant
Nos.~\href{https://doi.org/10.55776/COE1}{10.55776/COE1},
\href{https://doi.org/10.55776/F71}{10.55776/F71},
\href{https://doi.org/10.55776/P36305}{10.55776/P36305} and \href{https://doi.org/10.55776/ESP5750624}{10.55776/ESP5750624}), by the European Union
-- NextGenerationEU, and by the European Union’s Horizon 2020 research and
innovation program through Grant No.~863476 (ERC-CoG \mbox{SEQUAM}).
\end{acknowledgments}

\bibliography{literature}

\onecolumngrid
\appendix

\section{Transformation of Harmonic Oscillator Dynamics into a Schrödinger Equation} \label{sec:oscillators_to_se}

For completeness we repeat here the results of Ref.~\cite{Babbush_2023}: We recapitulate how to rephrase the harmonic oscillator equation \eqref{eq:hosc-pde} as a Schrödinger equation with a real-valued Hamiltonian, which then provides the starting point for the sign-split embedding used in \autoref{sec:ho<qw}, which ultimately leads to the sought-after quantum walk equation \eqref{eq:pde-randwalk-complex-c}.

We start with the construction of a matrix $B$ that satisfies $A=B^\dagger B$:
Given the weighted graph $G=(V,E)$ of the harmonic oscillator system, with
the edge weight of $(v,w)$ equal to $\kappa_{vw}$ (including self-edges), $B$
is an $|E|\times |V|$ matrix, i.e., it maps vertices to edges. For an
edge $e=(w,u)\in E$, with the convention that $w\le u$, and $v\in V$,
we define 
\begin{align}
B_{ev}&=\sqrt{\kappa_{wu}}\mbox{\ if\ } v=w \ , \nonumber
\\
B_{ev}&=-\sqrt{\kappa_{wu}}\mbox{\ if\ } v=u\ne w \ , \nonumber
\\
B_{ev}&=0\mbox{\ if\ } v\ne w,u \ . \label{eq:def-B}
\end{align}
It is immediate to see that with this definition, $B^\dagger
B=A$.\footnote{
For $(v,v')=e^*\in E$, 
$(B^\dagger B)_{vv'} = \sum_{e} B_{ev}B_{ev'} 
= B_{e^*v}B_{e^*v'} = -\kappa_{vv'}=A_{vv'}$; 
for $(v,v')\not\in E$, 
$(B^\dagger B)_{vv'} = \sum_{e} B_{ev}B_{ev'} =
0=A_{vv'}$;
and 
$(B^\dagger B)_{vv} = \sum_{w\ne v\,:\,e=(v,w)\in E}
|B_{ev}|^2 = \sum_w \kappa_{vw}=A_{vv}$.
}
Moreover, given oracle access to $\kappa$ (or equivalently, to $A$), we can
efficiently build an oracle for $B$ (assuming that edges are encoded as pairs of
vertices). Note that this requires a way to efficiently enumerate/index all
edges. This can for instance be achieved by looping over all vertices and, for
every vertex $w$, keeping a list of length $d$ (the maximum degree of the graph)
which enumerates the at most $d$ edges $(w,u)$ incident to it with $w\le u$,
with padding as needed; such a list can be efficiently generated on the fly
given oracle access to $\kappa$.\footnote{Alternatively, as done in
Ref.~\cite{Babbush_2023}, one can use all pairs $(v,w)$ to index the rows
of $B$,  padding rows corresponding to unused edges with zeros; this gives rise
to a larger matrix $B$ of size $N^2\times N$.}

Now let $c=\big({c^V\atop c^E}\big)$, where $c^V=(c^V_v)_{v\in
V}$ and $c^E=(c^E_e)_{e\in E}$ are complex vectors
indexed by vertices and edges, respectively. Consider the differential
equation
\begin{equation}
\begin{pmatrix} \dot c^E(t) \\ \dot c^V(t) \end{pmatrix}
=
-i
\underbrace{\begin{pmatrix} 0 & B \\ B^\dagger & 0
\end{pmatrix}}_{=:\tilde T}
\begin{pmatrix} c^E(t) \\ c^V(t) \end{pmatrix}\ .
\label{eq:hosc-to-qwalk-schroedingereq}
\end{equation}
This is a Schrödinger equation for $c=\big({c^V\atop c^E}\big)$, as 
$\tilde T$ is Hermitian by construction. If we assume that at some 
given time $t_0$, 
\begin{equation}
c^V(t_0)\in \mathbb R^{|V|}\mbox{\quad and\quad} 
c^E(t_0)\in i\, \mathrm{Im}_\mathbb{R}B\ ,
\label{eq:hosc-to-qwalk-subspace-constraint}
\end{equation}
with $\mathrm{Im}_\mathbb{R}B = \{Bc^V|c^V \in \mathbb R^{|V|}\}$ the real
image of $B$, 
then Eq.~\eqref{eq:hosc-to-qwalk-schroedingereq} tells us that 
\begin{align*}
\dot c^E(t_0) & = -iBc^V\in i\,\mathrm{Im}_\mathbb{R}B\ ,
\\
\dot c^V(t_0) & = -iB^\dagger c^E \in \mathbb R^{|V|}\ .
\end{align*}
This implies that the subspace given by the constraint
\eqref{eq:hosc-to-qwalk-subspace-constraint} is closed under the
evolution \eqref{eq:hosc-to-qwalk-schroedingereq}, and 
we will restrict ourselves to this closed subspace from now on
by choosing a corresponding initial condition at time $t_0=0$.

In this subspace, we can define a bijection
between $c=\big({c^E\atop c^V}\big)$ and the state $\big({q \atop p}\big)$
of the harmonic oscillator system by virtue of
\begin{equation}
\begin{pmatrix} c^E(t)\\c^V(t)\end{pmatrix}
=
\frac{1}{\sqrt{2\mathcal H}}
\begin{pmatrix} -iB q(t) \\ p(t) \end{pmatrix}\ ,
\label{eq:hosc-to-qwalk-substitution} 
\end{equation}
where we have used that $\mathcal H = \tfrac12 p(t)^Tp(t) + \tfrac12
q(t)^TB^\dagger Bq(t)$ is the conserved quantity in the Harmonic
Oscillator Problem in order to fix the normalization such that $|c^E(t)|^2+|c^V(t)|^2=1$. 
Substituting this into Eq.~\eqref{eq:hosc-to-qwalk-schroedingereq} and
multiplying the upper block by $-iB^\dagger$ yields the desired
differential equation \eqref{eq:hosc-pde} describing the dynamics of the
harmonic oscillator system. Given that we initialize our
system subject to the constraint
\eqref{eq:hosc-to-qwalk-subspace-constraint}, the system's dynamics under
Eq.~\eqref{eq:hosc-to-qwalk-schroedingereq} is thus in one-to-one
correspondence to that  of the harmonic oscillators.

\section{Interpretation of the Sign-Split Embedding}
\label{app:sign-separation}

In this section, we provide some examples of the mapping from a quantum walk graph to a harmonic oscillator system to demonstrate the sign-split embedding introduced in Section~\ref{sec:sign-trick}.

First, let us summarize the steps of the mapping: We start with a matrix $T$ that is the adjacency matrix of a quantum walk graph, i.e., $T$ is symmetric and entry-wise non-negative. We then calculate $\tilde{A}=T^2$, where $\tilde{A}$ defines an oscillator network (with unit masses) through $\kappa_{vw}=-\tilde{A}_{vw}$ for $v\neq w$ and $\kappa_{vv}=\sum_{w}\tilde{A}_{vw}$. Since the matrix $\tilde{A}$ is positive semi-definite by construction, the dynamics of this oscillator network are stable. Despite that, since $\tilde{A}$ is entry-wise non-negative, all springs between two masses have negative spring constants given by the off-diagonal entries of $-\tilde{A}$. The effect of those negative springs, which push the masses away from each other, is counteracted by the springs between the masses and the fixed wall. They have large positive spring constants given by the row or column-sums of $\tilde{A}$ and stabilize the system.

Given this $\tilde{A}$, we want another oscillator system $A$ that contains only springs with positive spring constants and has the same dynamics as $\tilde{A}$ (in a subspace of phase space that is invariant under time evolution). So we apply the sign-split embedding to $\tilde{A}$, which consists of a diagonal part $A^d$ and an off-diagonal part $A^o$:
\begin{align}
    &\text{real-symmetric }  && \text{sign-split} \nonumber\\
    &\tilde{A}=A^d+A^o && A= 
    \begin{pmatrix}
        A^d &-A^o \\
        -A^o &A^d
    \end{pmatrix}
    \ .
\end{align}
Applying the sign-split embedding yields a matrix $A$ that acts on a doubled system, which is interpreted as two copies of the original. When the displacements and momenta of the doubled system are restricted to the anti-symmetric subspace, where the masses in the two copies move in opposite directions, $\tilde{A}$ and $A$ produce the same dynamics.

The non-zero spring constants of $A$ can be obtained from $A^d$ and $A^o$ through $\kappa_{\sigma_1(v) \sigma_2(w)}=\kappa_{\sigma_2(v) \sigma_1(w)}=A^o_{vw}$, $\kappa_{\sigma_1(v) \sigma_1(v)}=\kappa_{\sigma_2(v) \sigma_2(v)}=A^d_{vv}-\sum_w A^o_{vw}$, where $\sigma_1$ and $\sigma_2$ map the indices of the original system to the indices of the first and second copy of the doubled system, respectively.

This will make all springs between masses positive, at the expense of reducing the spring constants to the wall, which can become negative. There are three cases:
\begin{itemize}
    \item[] case 1: all springs are positive, $A$ is positive semi-definite. (Example~\ref{ex:1}, Example~\ref{ex:4}).
    \item[] case 2: some springs to the wall are negative, $A$ is positive semi-definite.
    \item[] case 3: some or all springs to the wall are negative, $A$ is not positive semi-definite (Example~\ref{ex:3}).
\end{itemize}

In all three cases, if the oscillator system is initialized with an antisymmetric condition, its dynamics are stable. This is necessarily the case as it simulates a quantum walk on graph $T$.

We observe that we can always force the desired case 1 by adding a constant $\gamma$ to the diagonal of $T$ before squaring, which introduces a constant phase to the quantum walk amplitudes. This is demonstrated by Example~\ref{ex:3} and \ref{ex:4}.
(Note that this works only if $\tilde{A}$ is positive semi-definite. If this is not the case, the sign-split embedding can still be applied to $\tilde{A}$, resulting in positive springs between masses but negative springs to the wall in the doubled system. This will result in case 3, since $A$ cannot be positive semi-definite if $\tilde{A}$ is not.)

In the following figures, negative spring constants are indicated by red springs, and larger spring constants are indicated by closer windings of the coils:

\begin{example}
\label{ex:1}
We start from a graph consisting of three vertices. We will see that the result of the sign-split embedding falls in case 1. The corresponding graphs are shown in Fig.~\ref{fig:1}.
\begin{align}
    & T_1=
    \begin{pmatrix}
        0 &1 &0 \\
        1 &0 &1 \\
        0 &1 &0
    \end{pmatrix}
    && \tilde{A}_1=T_1^2=
    \begin{pmatrix}
        1 &0 &1 \\
        0 &2 &0 \\
        1 &0 &1
    \end{pmatrix}
    && A_1= 
    \begin{pmatrix}
        1 &0 &0  &0 &0 &-1 \\
        0 &2 &0  &0 &0 &0 \\
        0 &0 &1  &-1 &0 &0 \\
        0 &0 &-1 &1 &0 &0 \\
        0 &0 &0  &0 &2 &0 \\
        -1 &0 &0 &0 &0 &1
    \end{pmatrix}
\end{align}
\begin{figure}[h]
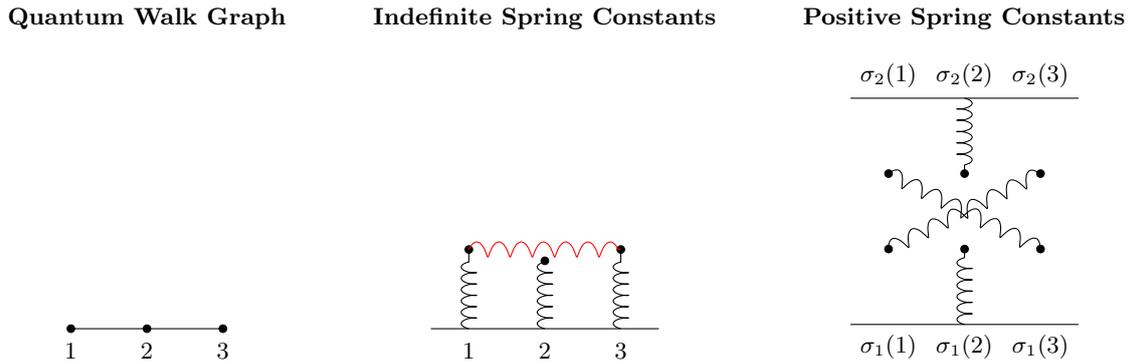

    \centering
    \begin{tabular*}{\textwidth}{c@{\extracolsep{\fill}}cccc}
        &\textbf{Quantum Walk Graph} & \textbf{Indefinite Spring Constants} & \textbf{Positive Spring Constants}& \\[3mm]
        &\ExampleIGraph & \ExampleIHO & \ExampleIHODouble&
    \end{tabular*}
    \caption{Left: Graph corresponding to $T_1$; Middle: System of unit masses and springs corresponding to $\tilde{A}_1=T_1^2$ with $\kappa_{11}=\kappa_{22}=\kappa_{33}=2$ and $\kappa_{13}=-1$, Right: System of unit masses and springs corresponding to $A_1$ with $\kappa_{\sigma_1(2)\sigma_1(2)}=2$ and $\kappa_{\sigma_1(1)\sigma_2(3)}=1$.}
    \label{fig:1}
\end{figure}
\end{example}

\begin{example}
\label{ex:3}
We start from a graph $T_2$ consisting of 4 vertices in a star arrangement and calculate $T_2^2=A_2^d+A_2^o$. After the sign-split embedding, the network $A$ has negative springs coupling to the wall (case 3). In general, this happens if the graph contains vertices with more next-nearest neighbors than neighbors (vertices 2, 3, and 4 in this example). The corresponding graphs are shown in Fig.~\ref{fig:2}.
\begin{align} 
    & T_2=
    \begin{pmatrix}
        0 &1 &1 &1\\
        1 &0 &0 &0 \\
        1 &0 &0 &0 \\
        1 &0 &0 &0 
    \end{pmatrix}
    & A^d_2= 
    \begin{pmatrix}
        3 &0 &0 &0 \\
        0 &1 &0 &0 \\
        0 &0 &1 &0 \\
        0 &0 &0 &1 \\
    \end{pmatrix}
    & \hspace{0.5cm} A^o_2= 
    \begin{pmatrix}
        0 &0 &0 &0 \\
        0 &0 &1 &1 \\
        0 &1 &0 &1 \\
        0 &1 &1 &0 \\
    \end{pmatrix}
\end{align}

\begin{figure}[h]
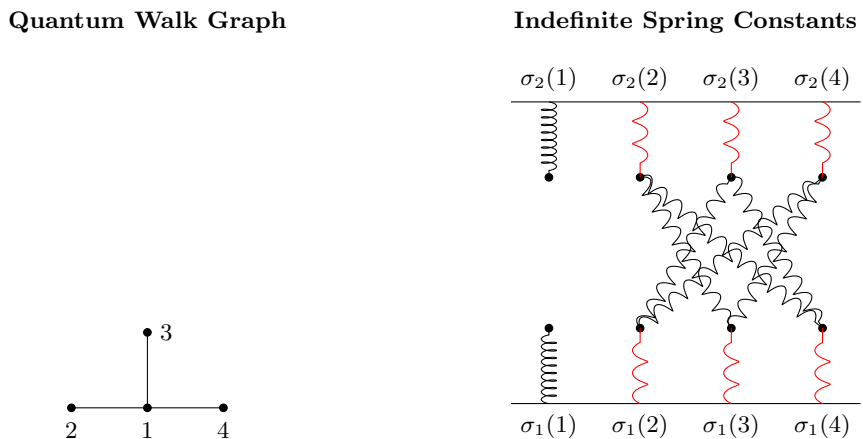
 
    \centering
    \begin{tabular*}{\textwidth}{c@{\extracolsep{\fill}}ccc}
        &\textbf{Quantum Walk Graph} & \textbf{Indefinite Spring Constants} & \\[3mm]
    &\ExampleIIIGraph & \ExampleIIIHODouble
    \end{tabular*}
    \caption{Left: Graph corresponding to $T_2$. Right: Doubling leads to negative spring couplings to the wall: $\kappa_{\sigma_1(2)\sigma_1(2)}=\kappa_{\sigma_1(3)\sigma_1(3)}=\kappa_{\sigma_1(4)\sigma_1(4)}=-1$, $\kappa_{\sigma_1(1)\sigma_1(1)}=3$, $\kappa_{\sigma_1(2)\sigma_2(3)}=\kappa_{\sigma_1(2)\sigma_2(4)}=\kappa_{\sigma_1(3)\sigma_2(4)}=1$.
    }
    \label{fig:2}
\end{figure}
\end{example}

\begin{example}
\label{ex:4}
By adding a sufficiently large diagonal, we can make $A$ positive definite using $T_3^2=A_3^d+A_3^o$ with $T_3=T_2+6 \mathds{1}$. The corresponding graphs are shown in Fig.~\ref{fig:3}. Here, coil drawings are replaced by straight lines with thickness indicating the coupling strength.
\begin{align}
    & T_3=T_2+6 \mathds{1}=
    \begin{pmatrix}
        6 &1 &1 &1 \\
        1 &6 &0 &0 \\
        1 &0 &6 &0 \\
        1 &0 &0 &6 
    \end{pmatrix}
    & A^d_3= 
    \begin{pmatrix}
        39 &0 &0 &0 \\
        0 &37 &0 &0 \\
        0 &0 &37 &0 \\
        0 &0 &0 &37
    \end{pmatrix}
     & \hspace{0.5cm} A^o_3= 
    \begin{pmatrix}
        0 &12 &12 &12 \\
        12 &0 &1 &1 \\
        12 &1 &0 &1 \\
        12 &1 &1 &0
    \end{pmatrix}
\end{align}
\begin{figure} [h]
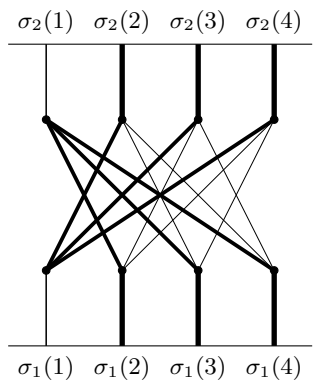
 
    \centering
    \begin{tabular*}{\textwidth}{c@{\extracolsep{\fill}}ccc}
        &\textbf{Quantum Walk Graph} & \textbf{Positive Spring Constants} & \\[3mm]
    &\ExampleIIIGraphDiagonal & \ExampleIIIHODiagonal & 
    \end{tabular*}
    \caption{Left: Graph $T_3$ constructed from $T_2$ by adding a weighted self-loop to every node. Right: For the doubled spring system we get $\kappa_{\sigma_1(2)\sigma_2(3)}=\kappa_{\sigma_1(2)\sigma_2(4)}=\kappa_{\sigma_1(3)\sigma_2(4)}=1$, $\kappa_{\sigma_1(1)\sigma_1(1)}=3$, $\kappa_{\sigma_1(1)\sigma_2(2)}=\kappa_{\sigma_1(1)\sigma_2(3)}=\kappa_{\sigma_1(1)\sigma_2(4)}=12$, and $\kappa_{\sigma_1(2)\sigma_1(2)}=\kappa_{\sigma_1(3)\sigma_1(3)}=\kappa_{\sigma_1(4)\sigma_1(4)}=23$. }
    \label{fig:3}
\end{figure}
\end{example}

\section{Mapping between Output Probability Distributions for the Quantum Walk to Harmonic Oscillator Mapping}
\label{app:distributions}
In the following, we derive the explicit form of the output probability distribution $P_{\rm QW}$ of the Quantum Walk Problem as a function of the output probabilities $P_{\rm HO}$ of the classical oscillators. We show that $P_{\rm QW}(v)$ can be calculated efficiently using $\mathcal{O}(d^3)$ queries to the adjacency matrix $T$. We can sample from a probability distribution of the form
\begin{equation}
P_{\mathrm{HO}}(s) = \left\{ \begin{array}{l@{\quad}l} 
	\tfrac12 p_v^2\ , & s=v\in V\\[1ex]
	\tfrac12 \kappa_{vw} (q_v-q_w)^2\ ,& s=(v,w)\in E \\[1ex]
	\tfrac12 \kappa_{vv} q_v^2\ ,& s=(v,v)\in E
\end{array}\right.\ ,
\end{equation}
which determines the quadratic observables via
\begin{align}
    p_v^2 &= 2P_{\mathrm{HO}}(v) \label{eq:pv2} \\
    q_v^2 &= \frac{2}{\kappa_{vv}} P_{\mathrm{HO}}((v,v)) \label{eq:qv2} \\
    q_v q_w &= \frac{P_{\mathrm{HO}}(( v, v))}{\kappa_{v v}} + \frac{P_{\mathrm{HO}}(( w, w))}{\kappa_{ww}} - \frac{P_{\mathrm{HO}}(( v, w))}{\kappa_{vw}} \ . \label{eq:qvqw}
\end{align}
The probability distribution of the output vertices of a quantum walk can be formulated as in Eq.~\eqref{eq:pQW-from-qq-pp}
\begin{align}
    P_{\mathrm{QW}}(v) & = |c_v|^2 = -\sum_{w\ne u} \tilde T_{vw}\tilde T_{vu} \bar q_{\sigma_1(w)} \bar q_{\sigma_2(u)} + \frac{1}{2} \sum_{w} \tilde T_{vw}^2 (\bar q_{\sigma_1(w)}^2 + \bar q_{\sigma_2(w)}^2) + \frac{1}{2} (\bar p_{\sigma_1(v)}^2 + \bar p_{\sigma_2(v)}^2) \ ,
    \label{eq:PQW}
\end{align}
with $\sigma_{1/2}$ mapping a vertex to the first or second copy respectively, as introduced in the main text. We can plug Eqs.~(\ref{eq:pv2} -- \ref{eq:qvqw}) into Eq.~\eqref{eq:PQW} and get
\begin{align}
    P_{\mathrm{QW}}(v) & = - \sum_{w\ne u} \tilde T_{vw}\tilde T_{vu} \left( \frac{P_{\mathrm{HO}}(( \sigma_1(w), \sigma_1(w)))}{\kappa_{\sigma_1(w)\sigma_1(w)}} + \frac{P_{\mathrm{HO}}((\sigma_2(u), \sigma_2(u)))}{\kappa_{\sigma_2(u) \sigma_2(u)}} - \frac{P_{\mathrm{HO}}((\sigma_1(w), \sigma_2(u)))}{\kappa_{\sigma_1(w)\sigma_2(u)}} \right) \nonumber \\
    &+ \sum_{w} \tilde T_{vw}^2\frac{P_{\mathrm{HO}}((\sigma_1(w), \sigma_1(w)))}{\kappa_{\sigma_1(w)\sigma_1(w)}} + \sum_{w} \tilde T_{vw}^2 \frac{P_{\mathrm{HO}}((\sigma_2(w), \sigma_2(w)))}{\kappa_{\sigma_2(w)\sigma_2(w)}} + P_{\mathrm{HO}}(\sigma_1(v)) + P_{\mathrm{HO}}(\sigma_2(v)) \ . \label{eq:P_QW(v)}
\end{align}
From here, we can read off the nonzero elements $C_{vs}$ such that $P_{\rm QW}(v) = \sum_s C_{vs} P_{\rm HO}(s)$ as
\begin{align}
    C_{v (\sigma_i(w), \sigma_i(w))} &= \frac{\tilde T_{vw}}{\kappa_{\sigma_i(w)\sigma_i(w)}} \left( \tilde T_{vw} - \sum_{u\ne w}\tilde T_{vu} \right) \label{eq:Cv_ww} \\
    C_{v (\sigma_1(w), \sigma_2(u))} &= \frac{\tilde T_{vu} \tilde T_{vw}}{\kappa_{\sigma_1(w) \sigma_2(u)}} \text{ for } w\neq u \label{eq:Cv_wu}
\end{align}
if $(v,w) \in E$ and $(v,u) \in E$; and for the vertex sample we get
\begin{align}
        C_{v \sigma_i(w)} &= \delta_{vw} \ . \label{eq:Cv_w}
\end{align}
In order to interpret the $C_{vs}$ as conditional probabilities, $C_{vs}\geq 0$ and $\sum_v C_{vs} = 1$ need to be satisfied. While $C_{v (\sigma_1(w), \sigma_2(u))}\geq 0$ and $C_{v \sigma_i(w)}\geq 0$ always hold (see Eqs.~\eqref{eq:Cv_wu} and \eqref{eq:Cv_w}, respectively), the coefficients corresponding to wall-springs $C_{v (\sigma_i(w), \sigma_i(w))}$ \eqref{eq:Cv_ww} can be negative. This means that we cannot directly map single samples of the harmonic oscillator system probabilistically to samples of the quantum walk using $C_{vs}$ as conditional probabilities.

To map the decision problems, we do the following: To calculate a single probability $P_{\mathrm{QW}}(v)$, $\mathcal{O}(d^2)$ probabilities $P_{\mathrm{HO}}(s)$ are needed to polynomial precision. Each of those is calculated by a binary search over the thresholds $a$ and $b$ of the harmonic oscillator decision problem, requiring $\mathcal{O}(\log n)$ adaptive calls to the decision oracle to reach $1/\poly(n)$ precision. This requires a total of $\mathcal{O}(d^2 \log n)$ calls to the decision oracle for the Harmonic Oscillator Problem. If $d=\poly(n)$, this is equivalent to $\poly(n)$ calls.

To evaluate $P_{\mathrm{QW}}(v)=\sum_s C_{vs} P_{\mathrm{HO}}(s)$ given by Eq. \eqref{eq:P_QW(v)}, we additionally require $\mathcal{O}(d^3)$ queries to the oracle for $T$:
This is due to the evaluation of the wall-spring constants $\kappa_{\sigma_{i}(w)\sigma_{i}(w)}=\tilde{A}_{ww}-\sum_{y\neq w}\tilde{A}_{wy}=\sum_x \tilde{T}_{wx}^2-\sum_{y\neq w} \sum_x \tilde{T}_{wx}\tilde{T}_{xy}$, where $w$ are all the neighbors of vertex $v$, and $y$ are all the next-nearest neighbors of $w$. In summary, we need to explore the graph to a distance 3 starting from vertex $v$ (following the path $v \rightarrow w \rightarrow x \rightarrow y$), which inherently requires $\mathcal{O}(d^3)$ queries to the oracle for $T$.

Observe that the normalization condition, $\sum_v C_{vs} = 1$, is satisfied. For $s = \sigma_i(v)$, this is trivially fulfilled. For $s=(\sigma_1(w), \sigma_2(u))$, one can use the relation $\tilde T^2 = \tilde A$ together with $\tilde A_{wu} = \kappa_{\sigma_1(w)\sigma_2(u)}$ for $w\neq u$ (recall Eq.~\eqref{eq:A_of_QW}), which yields $\sum_v C_{v (\sigma_1(w), \sigma_2(u))} = 1$. Finally, for $s= (\sigma_i(w), \sigma_i(w))$ we get
\begin{align}
    \sum_v C_{v(\sigma_i(w), \sigma_i(w))} &= \sum_v \frac{\tilde T_{vw}}{\kappa_{\sigma_i(w)\sigma_i(w)}} \left( \tilde T_{vw} - \sum_{u\ne w}\tilde T_{vu} \right) = \frac{1}{\kappa_{\sigma_i(w)\sigma_i(w)}} \left( \tilde A_{ww} - \sum_{u\ne w} \tilde A_{wu} \right) = 1 \ ,
\end{align}
from $\tilde A_{ww} = \kappa_{\sigma_i(w)\sigma_i(w)} + \sum_{u\neq w} \kappa_{\sigma_1(w)\sigma_2(u)}$. 

Note that the normalization condition $\sum_v C_{vs}=1$ implies that the
$C_{v(\sigma_1(w),\sigma_2(u))}$ in Eq.~\eqref{eq:Cv_wu} are bounded, as they
are all non-negative and sum to one; while the $C_{v(\sigma_i(w),\sigma_i(w))}$,
Eq.~\eqref{eq:Cv_ww}, are bounded due to the existence of a constant uniform
lower bound on the $\kappa_{\bar v\bar v}$, cf.\
Sec.~\ref{sec:qw-to-ho-dynamics}. Their boundedness is important to establish a
mapping from an oracle that returns estimates of $P_\mathrm{QW}$ with a $1/\poly(n)$
accuracy to one that estimates $P_\mathrm{HO}$ with a $1/\poly(n)$ accuracy.

\vspace{1em}

In order to get a better understanding for the necessity of negative $C_{vs}$, consider the following simple example.
\begin{example}
    Consider a quantum walk with two vertices and an edge in between described by the (self-loop corrected) adjacency matrix
    \begin{align}
        \tilde{T} = \begin{pmatrix}
        2 & 1 \\ 1 & 2
        \end{pmatrix} \qquad \implies \qquad \tilde{T}^2 = \begin{pmatrix}
            5 & 4 \\ 4 & 5
        \end{pmatrix} \ .
    \end{align}
    This implies $\kappa_{\sigma_1(1)\sigma_2(2)} = 4$ and $\kappa_{\sigma_i(1)\sigma_i(1)} = \kappa_{\sigma_i(2)\sigma_i(2)} = 1$. Calculating $C_{vs}$ with $v \in \{1,2\}$ by plugging into Eq.~\eqref{eq:Cv_ww} yields
    \begin{align}
        C_{1(\sigma_i(1), \sigma_i(1))} = 2, \qquad C_{1(\sigma_i(2), \sigma_i(2))} = -1, \qquad C_{2(\sigma_i(1), \sigma_i(1))} = -1, \qquad C_{2(\sigma_i(2), \sigma_i(2))} = 2,
    \end{align}
    as well as $C_{1 (\sigma_1(1), \sigma_2(2))} = C_{1 (\sigma_2(1), \sigma_1(2))} = \frac{1}{2}$. With this, we can calculate the probabilities of finding a quantum walker at vertex $1$ or $2$ as
    \begin{align}
        &P_{\mathrm{QW}}(1) = \sum_s C_{1s} P_{\rm HO} (s) = 4 P_{\rm HO}((\sigma_1(1), \sigma_1(1))) - 2 P_{\rm HO}((\sigma_1(2), \sigma_1(2))) + P_{\rm HO}((\sigma_1(1), \sigma_2(2))) \label{eq:PQWv1}\\
        &P_{\mathrm{QW}}(2) = \sum_s C_{2s} P_{\rm HO} (s) = 4 P_{\rm HO}((\sigma_1(2), \sigma_1(2))) - 2 P_{\rm HO}((\sigma_1(1), \sigma_1(1))) + P_{\rm HO}((\sigma_1(1), \sigma_2(2))),
        \label{eq:PQWv2}
    \end{align}
    where an overall factor of $2$ comes from the terms with $\sigma_1$ and $\sigma_2$ interchanged. Since the walker can only be at vertex $1$ or $2$, $P_{\mathrm{QW}}(1)+P_{\mathrm{QW}}(2)=1$ must hold. 
    
    Now consider that we measure the oscillator system at a time where only the masses $\sigma_i(1)$ are displaced, i.e., $p_{\sigma_i(1)} = p_{\sigma_i(2)} = 0$ and $q_{\sigma_i(2)} = 0$. The energy is then given by 
    \begin{align}
        E = 
        \kappa_{\sigma_i(1)\sigma_i(1)} \ q_{\sigma_i(1)}^2 + \kappa_{\sigma_1(1)\sigma_2(2)} \ q_{\sigma_i(1)}^2 = (1+4) \ q_{\sigma_i(1)}^2.
    \end{align}
    From $E=1$ follows that $q_{\sigma_i(1)}^2=\frac{1}{5}$, which gives the following probabilities for the harmonic oscillators:
    \begin{align}
        &P_{\rm HO}((\sigma_1(1), \sigma_1(1))) = \frac{1}{10}\\
        &P_{\rm HO}((\sigma_1(2), \sigma_1(2))) = 0 \\
        &P_{\rm HO}((\sigma_1(1), \sigma_2(2))) = \frac{2}{5}.
    \end{align}
    Inserting into Eq.~\eqref{eq:PQWv1} and \eqref{eq:PQWv2} yields 
    \begin{align}
        &P_{\mathrm{QW}}(1) = \frac{2}{5} - 0 + \frac{2}{5} = \frac{4}{5} \\
        &P_{\mathrm{QW}}(2) = 0 - \frac{1}{5} + \frac{2}{5} = \frac{1}{5}.
    \end{align}

    We observe that the term $P_{\rm HO}((\sigma_1(1), \sigma_2(2)))$ contributes equally to $P_{\mathrm{QW}}(1)$ and $P_{\mathrm{QW}}(2)$. Since $P_{\mathrm{QW}}(1)$ already has the value $\frac{4}{5}$ and $P_{\mathrm{QW}}(2)$ must also include $P_{\rm HO}((\sigma_1(1), \sigma_2(2))) = \frac{2}{5}$, one can see that the negative term is necessary in order to normalize the probabilities, that is $P_{\mathrm{QW}}(1)+P_{\mathrm{QW}}(2)=1$.
\end{example}

\section{Simulating Laplacian quantum walks by harmonic oscillators}
\label{app:laplacian}

Instead of the adjacency matrix $T$, one can also define the quantum walk via the Laplacian $L$:
\begin{equation} 
    \dot{c} = -iLc \ ,
\end{equation}
in analogy with Eq.~\eqref{eq:pde-randwalk-complex-c}. The Laplacian is defined as $L=D-T$, where $T$ is again the adjacency matrix and $D$ is the degree matrix. To reduce this quantum walk to a harmonic oscillator system, we proceed similarly as before. First, we define a matrix $\tilde{L}:=D-T+\gamma \mathds{1}$, which is then squared to give $\tilde{A}:=\tilde{L}^2$. We use the sign-split embedding on $\tilde{A}$ and define $A$ in terms of $T$ and $D$ as
\begin{equation}
    A:=
    \begin{pmatrix}
        (D+\gamma \mathds{1})^2+(T^2)_d -DT -TD -2 \gamma T & -(T^2)_o\\
        -(T^2)_o & (D+\gamma \mathds{1})^2+(T^2)_d -DT -TD -2 \gamma T
    \end{pmatrix}\ , \label{eq:A_of_QW_Laplacian}
\end{equation} 
where $(T^2)_d$ and $(T^2)_o$ are the diagonal and off-diagonal parts of $T^2$, respectively. Without loss of generality, we assume that the diagonal of $T$ is zero, since the Laplacian is unaffected by self-loops. Simulating a query to the oracle for $A$ requires $\mathcal{O}(d^2)$ queries to the oracle for $L$.

To verify that this choice of $A$ results in wall-springs $\kappa_{\sigma_i(v) \sigma_i(v)}$ that are lower-bounded by a constant, we start by writing out the sum explicitly: We have
\begin{align}
    \kappa_{\sigma_i(v) \sigma_i(v)} &= \sum_{u} \left( A_{\sigma_i(v) \sigma_1(u)} + A_{\sigma_i(v) \sigma_2(u)} \right) \nonumber \\
    &= (D_{vv} + \gamma )^2 + (T^2)_{vv} - D_{vv} \sum_u T_{vu} - \sum_u T_{vu}D_{uu} - 2 \gamma \sum_u T_{vu} + (T^2)_{vv} - \sum_u (T^2)_{vu} \ , 
\end{align}
where we have used $-(T^2)_o=(T^2)_d-T^2$. We now use $D_{vv}=\sum_u T_{vu}$ and, consequently, $\sum_u (T^2)_{vu}=\sum_u T_{vu}D_{uu}$, and get
\begin{align}
    \kappa_{\sigma_i(v) \sigma_i(v)} &= (D_{vv} + \gamma )^2 + (T^2)_{vv} - D_{vv}^2 - \sum_u T_{vu}D_{uu} - 2 \gamma D_{vv} + (T^2)_{vv} - \sum_u T_{vu}D_{uu} \nonumber \\
    &= \gamma^2 + 2 (T^2)_{vv} - 2 \sum_u T_{vu}D_{uu} \nonumber \\
    &\geq  \gamma^2 - 2d^2 \ . \label{eq:wall-springs_Laplacian}
\end{align}
We thus arrive at $\kappa_{vv} \geq 7d^2$ for $\gamma=3d$. This choice of $\gamma$ also ensures invertibility of $\tilde{L}$.

The initial conditions for the harmonic oscillator network given by Eq.~\eqref{eq:qp-initial-conditions} remain unchanged. For the read-out we have
\begin{equation}
\begin{aligned}
P_{\mathrm{QW}}(v) & = |c_v|^2 = a_v^2 + b_v^2 
\\ & =  ((\tilde Lq)_v)^2 + p_v^2
\\ & = \sum_{w,u} \tilde L_{vw} \tilde L_{vu} q_w q_u + p_v^2
\\ & = \frac{1}{2} \sum_{w} \tilde L_{vw}^2 \left( \bar q_{\sigma_1(w)}^2 + \bar q_{\sigma_2(w)}^2 \right) + \frac{1}{2} \left( \bar p_{\sigma_1(v)}^2 + \bar p_{\sigma_2(v)}^2 \right)
\\ & \qquad + \sum_{w (\ne v)} \tilde L_{vv}\tilde L_{vw} \left( \bar q_{\sigma_1(v)} \bar q_{\sigma_1(w)} + \bar q_{\sigma_2(v)} \bar q_{\sigma_2(w)} \right)
\\ & \qquad - \sum_{w\neq u \ (w,u\neq v)} \tilde L_{vw}\tilde L_{vu}\bar q_{\sigma_1(w)} \bar q_{\sigma_2(u)} \ .
\end{aligned}
\end{equation}
This choice is symmetric in both copies and again picks pairs $\bar q_{\bar w} \bar q_{\bar u}$ for which $(\bar w,\bar u)\in \bar E$.

We get
\begin{align}
     P_{\mathrm{QW}}(v) & = \sum_{w} \tilde L_{vw}^2 \sum_{i} \frac{P_{\mathrm{HO}}((\sigma_i(w), \sigma_i(w)))}{\kappa_{\sigma_i(w)\sigma_i(w)}} + \sum_i P_{\mathrm{HO}}(\sigma_i(v)) \nonumber \\
   & + \sum_{w (\ne v)} \tilde L_{vv}\tilde L_{vw} \sum_i \left( \frac{P_{\mathrm{HO}}(( \sigma_i(v), \sigma_i(v)))}{\kappa_{\sigma_i(v)\sigma_i(v)}} + \frac{P_{\mathrm{HO}}(( \sigma_i(w), \sigma_i(w)))}{\kappa_{\sigma_i(w)\sigma_i(w)}} - \frac{P_{\mathrm{HO}}(( \sigma_i(v), \sigma_i(w)))}{\kappa_{\sigma_i(v)\sigma_i(w)}} \right) \nonumber \\
   & - \sum_{w\ne u \ (w,u \neq v)} \tilde L_{vw}\tilde L_{vu} \left( \sum_i \frac{P_{\mathrm{HO}}(( \sigma_i(w), \sigma_i(w)))}{\kappa_{\sigma_i(w)\sigma_i(w)}} - \frac{P_{\mathrm{HO}}(( \sigma_1(w), \sigma_2(u)))}{\kappa_{\sigma_1(w)\sigma_2(u)}} \right)
   \ .
\end{align}

From here, we can read off the nonzero elements $C_{vs}$ such that $P_{\rm QW}(v) = \sum_s C_{vs} P_{\rm HO}(s)$ as
\begin{align}
    C_{v (\sigma_i(v), \sigma_i(v))} &= \frac{\tilde L_{vv}}{\kappa_{\sigma_i(v)\sigma_i(v)}} \left( \tilde L_{vv} + \sum_{w (\ne v)}\tilde L_{vw} \right) = \frac{\gamma \tilde L_{vv}}{\kappa_{\sigma_i(v)\sigma_i(v)}} \label{eq:Cv_vv_Laplacian} \\
    C_{v (\sigma_i(w), \sigma_i(w))} &= \frac{\tilde L_{vw}}{\kappa_{\sigma_i(w)\sigma_i(w)}} \left( \tilde L_{vv} + \tilde L_{vw} - \sum_{u (\ne v,w)}\tilde L_{vu} \right) = \frac{\tilde L_{vw}}{\kappa_{\sigma_i(w)\sigma_i(w)}} \left( 2 \tilde L_{vv} + 2 \tilde L_{vw} - \gamma \right) \text{ for } w\neq v \label{eq:Cv_ww_Laplacian} \\
    C_{v (\sigma_i(v), \sigma_i(w))} &= - \frac{\tilde L_{vv} \tilde L_{vw}}{\kappa_{\sigma_i(v) \sigma_i(w)}} \text{ for } w\neq v \label{eq:Cv_vw_Laplacian} \\
    C_{v (\sigma_1(w), \sigma_2(u))} &= \frac{\tilde L_{vw} \tilde L_{vu}}{\kappa_{\sigma_1(w) \sigma_2(u)}} \text{ for } w\neq u, \ w,u\neq v \label{eq:Cv_wu_Laplacian} \ ,
\end{align}
if $(v,w) \in E$ and $(v,u) \in E$; and where we have used that $\sum_{w}\tilde L_{vw}=\gamma$.
For the vertex sample we get
\begin{align}
    C_{v \sigma_i(w)} &= \delta_{vw} \ . \label{eq:Cv_w_Laplacian}
\end{align}

We now verify the normalization condition $\sum_v C_{vs}=1$ for each of those coefficients. We start with $s=(\sigma_i(w),\sigma_i(w))$ and insert Eqs.~\eqref{eq:Cv_vv_Laplacian} and \eqref{eq:Cv_ww_Laplacian}:
\begin{align}
\sum_v C_{v (\sigma_i(w), \sigma_i(w))} &=  C_{w (\sigma_i(w), \sigma_i(w))} + \sum_{v(\neq w)} C_{v (\sigma_i(w), \sigma_i(w))} \nonumber \\
&= \frac{1}{\kappa_{\sigma_i(w)\sigma_i(w)}} \left( \gamma \tilde L_{ww} + \sum_{v(\neq w)} \tilde L_{vw} \left( 2 \tilde L_{vv} + 2 \tilde L_{vw} - \gamma \right) \right) \nonumber \\
&= \frac{1}{\kappa_{\sigma_i(w)\sigma_i(w)}} \left( \gamma^2 + \gamma D_{ww} - \sum_{v} \left( \gamma T_{vw} + 2 D_{vv} T_{vw} - 2 T_{vw}^2 \right) \right) \nonumber \\
&= \frac{1}{\kappa_{\sigma_i(w)\sigma_i(w)}} \left( \gamma^2 - 2 \sum_{v} D_{vv} T_{vw} + 2 (T^2)_{ww} \right) = 1 \ ,
\end{align}
where a comparison with Eq.~\eqref{eq:wall-springs_Laplacian} lets us conclude the last equality.

We continue with $s=(\sigma_i(w),\sigma_i(u))$ for $w\neq u$. The only non-zero coefficients $C_{vs}$ are for $v=w$ and $v=u$ and are given by Eq.~\eqref{eq:Cv_vw_Laplacian}:
\begin{align}
\sum_v C_{v (\sigma_i(w), \sigma_i(u))} &=  C_{w (\sigma_i(w), \sigma_i(u))} + C_{u (\sigma_i(w), \sigma_i(u))} \nonumber \\
&= \frac{1}{\kappa_{\sigma_i(w)\sigma_i(u)}} \left( - \tilde L_{ww}\tilde L_{wu} - \tilde L_{uu}\tilde L_{uw} \right) \nonumber \\
&= \frac{T_{wu}}{\kappa_{\sigma_i(w)\sigma_i(u)}} \left( D_{ww} + D_{uu} + 2 \gamma \right) \nonumber \\
&= \frac{(DT + TD + 2 \gamma T)_{wu}}{\kappa_{\sigma_i(w)\sigma_i(u)}} = -\frac{A_{\sigma_i(w)\sigma_i(u)}}{\kappa_{\sigma_i(w)\sigma_i(u)}}  = 1 \ ,
\end{align}
where we have used the definition $A$ given by Eq.~\eqref{eq:A_of_QW_Laplacian}.

We continue with $s=(\sigma_1(w),\sigma_2(u))$ for $w\neq u$, where $C_{vs}$ is non-zero only if $v\neq w$ and $v\neq u$ and is given by Eq.~\eqref{eq:Cv_wu_Laplacian}:
\begin{align}
\sum_v C_{v (\sigma_1(w), \sigma_2(u))} &= \sum_{v(\neq w,u)} C_{v (\sigma_1(w), \sigma_2(u))} \nonumber \\
&= \frac{1}{\kappa_{\sigma_1(w)\sigma_2(u)}} \sum_{v(\neq w,u)} \tilde L_{vw}\tilde L_{vu} \nonumber \\
&= \frac{1}{\kappa_{\sigma_1(w)\sigma_2(u)}} \sum_{v(\neq w,u)} T_{vw} T_{vu} \nonumber \\
&= \frac{(T^2)_{wu}}{\kappa_{\sigma_1(w)\sigma_2(u)}} = -\frac{A_{\sigma_1(w)\sigma_2(u)}}{\kappa_{\sigma_1(w)\sigma_2(u)}} = 1 \ ,
\end{align}
where we have used again the definition $A$ given by Eq.~\eqref{eq:A_of_QW_Laplacian}.

Finally, the normalization condition is trivially fulfilled for $s=\sigma_i(w)$ with $C_{vs}$ given by Eq.~\eqref{eq:Cv_w_Laplacian}.

\end{document}

%% file: packages.tex
\usepackage{graphicx} 
\usepackage[a4paper, total={170mm, 237mm}]{geometry}
\usepackage[dvipsnames]{xcolor}
\usepackage[breaklinks=true]{hyperref}
\hypersetup{
    colorlinks = true,
    allcolors = RoyalPurple
}
\usepackage{amsfonts,amsmath,amssymb,amsthm}
\usepackage{marvosym}
\usepackage{dsfont}
\usepackage{braket}
\usepackage{tikz}
\usetikzlibrary{decorations.pathmorphing,patterns}
\usepackage{tikz-cd}
\usepackage{complexity}
\usepackage{orcidlink}
\usepackage{makecell}
\usepackage{booktabs}

\theoremstyle{plain} \newtheorem{theorem}{Theorem}
\theoremstyle{plain} 

%% file: definitions.tex
\DeclareMathOperator{\Mat}{Mat}

\newlang{\Hit}{QWHit}
\newlang{\Hita}{QWHitA}
\newlang{\Osc}{HOsc}
\newlang{\Osca}{HOscA}

\newtheorem{definition}{Definition}[]

\newtheorem{example}[definition]{Example}

\renewcommand{\ket}[1]{\vert #1 \rangle}
\renewcommand{\bra}[1]{\langle #1 \vert}

\newcommand{\abs}[1]{\left\lvert #1 \right\rvert}

\renewcommand{\epsilon}{\varepsilon}

\newcommand{\ExampleIGraph}{
\scalebox{1}{
\begin{tikzpicture}
    \filldraw (0,0.5) circle (.05cm);
    \filldraw (1,0.5) circle (.05cm);
    \filldraw (2,0.5) circle (.05cm);

    \node at (0,0.2) {1};
    \node at (1,0.2) {2};
    \node at (2,0.2) {3};

    \draw (0,0.5) -- (2,0.5);
\end{tikzpicture}
    }
}

\newcommand{\ExampleIHO}{
\scalebox{1}{
\begin{tikzpicture}

    \filldraw (0,1.05) circle (.05cm);
    \filldraw (1,0.9) circle (.05cm);
    \filldraw (2,1.05) circle (.05cm);

    \node at (0,-0.3) {1};
    \node at (1,-0.3) {2};
    \node at (2,-0.3) {3};

    \draw (-0.5,0) -- (2.5,0);

    \draw[decoration={aspect=0.3, segment length=1.5mm, amplitude=1mm,coil},decorate] (0,0) -- (0,1.03);
    \draw[decoration={aspect=0.3, segment length=1.5mm, amplitude=1mm,coil},decorate] (1,0) -- (1,0.9);
    \draw[decoration={aspect=0.3, segment length=1.5mm, amplitude=1mm,coil},decorate] (2,0) -- (2,1.03);
    \draw[red, decoration={aspect=0.3, segment length=2.93mm, amplitude=1mm,coil},decorate] (0,1.05) -- (2,1.05);
\end{tikzpicture}
    }
}

\newcommand{\ExampleIHODouble}{
\scalebox{1}{
\begin{tikzpicture}
    \filldraw (0,1) circle (.05cm);
    \filldraw (1,1) circle (.05cm);
    \filldraw (2,1) circle (.05cm);

    \node at (0,-0.3) {$\sigma_1(1)$};
    \node at (1,-0.3) {$\sigma_1(2)$};
    \node at (2,-0.3) {$\sigma_1(3)$};

    \filldraw (0,2) circle (.05cm);
    \filldraw (1,2) circle (.05cm);
    \filldraw (2,2) circle (.05cm);

    \node at (0,3.3) {$\sigma_2(1)$};
    \node at (1,3.3) {$\sigma_2(2)$};
    \node at (2,3.3) {$\sigma_2(3)$};

    \draw (-0.5,0) -- (2.5,0);
    \draw (-0.5,3) -- (2.5,3);

    \draw[decoration={aspect=0.3, segment length=1.5mm, amplitude=1mm,coil},decorate] (1,0) -- (1,1);
    \draw[decoration={aspect=0.3, segment length=1.5mm, amplitude=1mm,coil},decorate] (1,3) -- (1,2);
    \draw[decoration={aspect=0.3, segment length=2.89mm, amplitude=1mm,coil},decorate] (0,1) -- (2,2);
    \draw[decoration={aspect=0.3, segment length=2.89mm, amplitude=1mm,coil},decorate] (0,2) -- (2,1);
\end{tikzpicture}
    }
}

\newcommand{\ExampleIIIGraph}{
\scalebox{1}{
\begin{tikzpicture}
    \filldraw (0,0.5) circle (.05cm);
    \filldraw (1,0.5) circle (.05cm);
    \filldraw (2,0.5) circle (.05cm);
    \filldraw (1,1.5) circle (.05cm);
    \node at (1,0.2) {1};    
    \node at (0,0.2) {2};
    \node at (2,0.2) {4};
    \node at (1.25,1.5) {3};

    \draw (0,0.5) -- (2,0.5);
    \draw (1,0.5) -- (1,1.5);
\end{tikzpicture}
    }
}

\newcommand{\ExampleIIIHODouble}{
\scalebox{1}{
\begin{tikzpicture}

    \filldraw (0.6,1) circle (.05cm);
    \filldraw (-0.6,1) circle (.05cm);
    \filldraw (1.8,1) circle (.05cm);
    \filldraw (3,1) circle (.05cm);
    \filldraw (0.6,3) circle (.05cm);
    \filldraw (-0.6,3) circle (.05cm);
    \filldraw (1.8,3) circle (.05cm);
    \filldraw (3,3) circle (.05cm);

    \node at (0.6,-0.3) {$\sigma_1(2)$};
    \node at (-0.6,-0.3) {$\sigma_1(1)$};
    \node at (1.8,-0.3) {$\sigma_1(3)$};
    \node at (3,-0.3) {$\sigma_1(4)$};
    \node at (0.6,4.3) {$\sigma_2(2)$};
    \node at (-0.6,4.3) {$\sigma_2(1)$};
    \node at (1.8,4.3) {$\sigma_2(3)$};
    \node at (3,4.3) {$\sigma_2(4)$};

    \draw (-1.1,0) -- (3.5,0);
    \draw (-1.1,4) -- (3.5,4);

    \draw[decoration={aspect=0.3, segment length=2.55mm, amplitude=1mm,coil},decorate] (0.6,1) -- (1.8,3);
    \draw[decoration={aspect=0.3, segment length=2.55mm, amplitude=1mm,coil},decorate] (0.6,3) -- (1.8,1);
    \draw[decoration={aspect=0.3, segment length=2.55mm, amplitude=1mm,coil},decorate] (0.6,1) -- (3,3);
    \draw[decoration={aspect=0.3, segment length=2.55mm, amplitude=1mm,coil},decorate] (0.6,3) -- (3,1);
    \draw[decoration={aspect=0.3, segment length=2.55mm, amplitude=1mm,coil},decorate] (1.8,1) -- (3,3);
    \draw[decoration={aspect=0.3, segment length=2.55mm, amplitude=1mm,coil},decorate] (1.8,3) -- (3,1);
    
    \draw[red, decoration={aspect=0.3, segment length=3mm, amplitude=1mm,coil},decorate] (0.6,4) -- (0.6,3);
    \draw[decoration={aspect=0.3, segment length=1mm, amplitude=1mm,coil},decorate] (-0.6,4) -- (-0.6,3);
    \draw[red, decoration={aspect=0.3, segment length=3mm, amplitude=1mm,coil},decorate] (1.8,4) -- (1.8,3);
    \draw[red, decoration={aspect=0.3, segment length=3mm, amplitude=1mm,coil},decorate] (3,4) -- (3,3);
    \draw[red, decoration={aspect=0.3, segment length=3mm, amplitude=1mm,coil},decorate] (0.6,0) -- (0.6,1);
    \draw[decoration={aspect=0.3, segment length=1mm, amplitude=1mm,coil},decorate] (-0.6,0) -- (-0.6,1);
    \draw[red, decoration={aspect=0.3, segment length=3mm, amplitude=1mm,coil},decorate] (1.8,0) -- (1.8,1);
    \draw[red, decoration={aspect=0.3, segment length=3mm, amplitude=1mm,coil},decorate] (3,0) -- (3,1);
\end{tikzpicture}
    }
}

\newcommand{\ExampleIIIGraphDiagonal}{
\scalebox{1}{
\begin{tikzpicture}
    \draw (0,0.5) node[fill,circle,scale=0.4](2) {};
    \draw (1,0.5) node[fill,circle,scale=0.4](1) {};
    \draw (2,0.5) node[fill,circle,scale=0.4](4) {};
    \draw (1,1.5) node[fill,circle,scale=0.4](3) {};
    
    \node at (0,0.75) {2};
    \node at (1.2,0.75) {1};
    \node at (2,0.75) {4};
    \node at (1.25,1.5) {3};

    \draw (0,0.5) -- (2,0.5);
    \draw (1,0.5) -- (1,1.5);

    \path (2) edge [in=240, out=300, looseness=30, line width=0.4mm] (2);
    \path (1) edge [in=240, out=300, looseness=30, line width=0.4mm] (1);
    \path (3) edge [in=60, out=120, looseness=30,  line width=0.4mm] (3);
    \path (4) edge [in=240, out=300, looseness=30, line width=0.4mm] (4);
    
\end{tikzpicture}
    }
}

\newcommand{\ExampleIIIHODiagonal}{
\scalebox{1}{
\begin{tikzpicture}

    \filldraw (0,1) circle (.05cm);
    \filldraw (1,1) circle (.05cm);
    \filldraw (2,1) circle (.05cm);
    \filldraw (3,1) circle (.05cm);
    \filldraw (0,3) circle (.05cm);
    \filldraw (1,3) circle (.05cm);
    \filldraw (2,3) circle (.05cm);
    \filldraw (3,3) circle (.05cm);

    \node at (0,-0.3) {$\sigma_1(1)$};     
    \node at (1,-0.3) {$\sigma_1(2)$};    
    \node at (2,-0.3) {$\sigma_1(3)$};
    \node at (3,-0.3) {$\sigma_1(4)$};
    \node at (0,4.3) {$\sigma_2(1)$};
    \node at (1,4.3) {$\sigma_2(2)$};
    \node at (2,4.3) {$\sigma_2(3)$};
    \node at (3,4.3) {$\sigma_2(4)$};

    \draw (-0.5,0) -- (3.5,0);
    \draw (-0.5,4) -- (3.5,4);

    \draw[line width=0.5mm] (1,1) -- (0,3);
    \draw[line width=0.5mm] (1,3) -- (0,1);
    \draw[line width=0.1mm] (1,1) -- (2,3);
    \draw[line width=0.1mm] (1,3) -- (2,1);
    \draw[line width=0.1mm] (1,1) -- (3,3);
    \draw[line width=0.1mm] (1,3) -- (3,1);
    \draw[line width=0.5mm] (0,1) -- (2,3);
    \draw[line width=0.5mm] (0,3) -- (2,1);
    \draw[line width=0.5mm] (0,1) -- (3,3);
    \draw[line width=0.5mm] (0,3) -- (3,1);
    \draw[line width=0.1mm] (2,1) -- (3,3);
    \draw[line width=0.1mm] (2,3) -- (3,1);
    
    \draw[line width=0.7mm] (1,4) -- (1,3);
    \draw[line width=0.2mm] (0,4) -- (0,3);
    \draw[line width=0.7mm] (2,4) -- (2,3);
    \draw[line width=0.7mm] (3,4) -- (3,3);
    \draw[line width=0.7mm] (1,0) -- (1,1);
    \draw[line width=0.2mm] (0,0) -- (0,1);
    \draw[line width=0.7mm] (2,0) -- (2,1);
    \draw[line width=0.7mm] (3,0) -- (3,1);
    
\end{tikzpicture}
    }
}